\documentclass[a4paper,10pt, epsfig]{article}
\def\letter{0}\def\pr{0}
\usepackage{epsfig}
\usepackage{epstopdf}
\usepackage{graphicx}
\usepackage{ifthen}
\usepackage{amsmath}
\usepackage[psamsfonts]{amssymb}
\usepackage{euscript}

\usepackage{latexsym}
\usepackage[arrow,matrix,curve]{xy}

\newskip\humongous \humongous=0pt plus 1000pt minus 1000pt

\newif\ifdtup

\def\,{\hspace{-.1cm}}
\def\hsp{,\hspace{.7cm}}

\def\fc#1#2 {\frac{n}{q}#1\frac{n}{q}#2}

\def\tg{{\tilde{g}}}

\newcommand{\vac}{\ensuremath{|0\rangle}}

\renewcommand{\sin}{\textrm{sin}}

\newcommand{\sech}{\textrm{sech}}

\def\exp#1{\hbox{\rm exp}\left(#1\right)}

\renewcommand{\theequation}{\arabic{section}.\arabic{equation}}
\renewcommand{\(}{\begin{equation}}
\renewcommand{\)}{end{equation} \vspace{-.05in}\linebreak}

\newcounter{saveeqn}
\newcounter{savealpheqn}

\newcommand{\alpheqn}{\setcounter{saveeqn}{\value{equation}}%
  \stepcounter{saveeqn}\setcounter{equation}{0}%
  \renewcommand{\theequation}{\mbox{\arabic{section}.\arabic{saveeqn}
\alph{equation}}}
  \renewcommand{\)}{\end{equation}}}
\def\part#1{\frac{\partial}{\partial{#1}}}%
\def\group#1{\refstepcounter{equation}\setcounter{saveeqn}
 {\value{equation}}%
  \label{#1}\setcounter{equation}{0}%
\renewcommand{\theequation}{\mbox{\arabic{section}.\arabic{saveeqn}
\alph{equation}}}
  \renewcommand{\)}{\end{equation}}}
\newcommand{\reseteqn}{\setcounter{equation}{\value{saveeqn}}%
  \renewcommand{\theequation}{\arabic{section}.\arabic{equation}}%
  \renewcommand{\)}{\end{equation}}}

\newcommand{\aalpheqn}{\setcounter{saveeqn}{\value{equation}}%
  \stepcounter{saveeqn}\setcounter{equation}{0}%
  \renewcommand{\theequation}{\mbox{
        \Alph{subsection}.\arabic{saveeqn}\alph{equation}}}
   \renewcommand{\)}{\end{equation}}}
\newcommand{\areseteqn}{\setcounter{equation}{\value{saveeqn}}%
  \renewcommand{\theequation}{\Alph{subsection}.\arabic{equation}}%
  \renewcommand{\)}{\end{equation}}}

\renewcommand{\thefootnote}{\alph{footnote}}
\renewcommand{\(}{\begin{equation}}
\renewcommand{\)}{\end{equation}}
\newcommand{\ba}{\begin{eqnarray}}
\newcommand{\ea}{\end{eqnarray}}
\newcommand{\cbp}{\mathop{\vtop{\ialign{##\crcr
   $\hfil\displaystyle{}\hfil$\crcr\noalign{\kern-13pt\nointerlineskip}
   \BIG{)}\hskip0pt\crcr\noalign{\kern3pt}}}}}
\newcommand{\pa}{\mathop{\vtop{\ialign{##\crcr

$\hfil\displaystyle{\oplus}\hfil$\crcr\noalign{\kern+1pt\nointerlineskip
}
   \hspace{.08in}$^{\alpha=0}$\hskip6pt\crcr\noalign{\kern3pt}}}}}
\renewcommand{\hsp}{,\hspace{.3in}}
\newcommand{\p}{^\prime}

\catcode`\@=11
\def\vereq#1#2{\lower3pt\vbox{\baselineskip1.5pt \lineskip1.5pt
\ialign{$\m@th#1\hfill##\hfil$\crcr#2\crcr\sim\crcr}}}
\catcode`\@=12

\renewcommand{\(}{\begin{equation}}
\renewcommand{\)}{\end{equation}}

\newcommand{\wvname}{\rho}

\def\pin#1{\int \frac{d#1}{2\pi}}
\def\pink#1{\int \frac{d^{#1}k}{(2\pi)^{#1}}}

\def\Bd#1{B^\dag_{k_{#1}}}

\def\df{\mathcal{D}_{\Phi_{x_0}}}

\newcommand{\beas}{\begin{eqnarray*}}
\newcommand{\eeas}{\end{eqnarray*}}

\newcommand{\bquo}{\begin{quote}}
\newcommand{\enqu}{\end{quote}}

\newcommand{\C}{{\mathbb C}}

\newcommand{\R}{{\mathbb R}}

\def\ch{{\mathcal{H}}}

\def\os{\omega_S}
\def\op{\omega_p}
\def\oq{\omega_q}
\def\ok#1{\omega_{k_{#1}}}

\def\v#1{V^{(#1)}(\Phi(x),x)}

\newcommand{\beq}{\begin{equation}}
\newcommand{\eeq}{\end{equation}}
\newcommand{\bea}{\begin{eqnarray}}
\newcommand{\eea}{\end{eqnarray}}

\newskip\humongous \humongous=0pt plus 1000pt minus 1000pt

\newif\ifdtup

\catcode`\@=11

\ifthenelse{\equal{\letter}{0}}{ %se lettere=0, commenta questo

\@addtoreset{equation}{section}
\def\theequation{\arabic{section}.\arabic{equation}}

\def\@normalsize{\@setsize\normalsize{15pt}\xiipt\@xiipt
\abovedisplayskip 14pt plus3pt minus3pt%
\belowdisplayskip \abovedisplayskip
\abovedisplayshortskip \z@ plus3pt%
\belowdisplayshortskip 7pt plus3.5pt minus0pt}

\def\small{\@setsize\small{13.6pt}\xipt\@xipt
\abovedisplayskip 13pt plus3pt minus3pt%
\belowdisplayskip \abovedisplayskip
\abovedisplayshortskip \z@ plus3pt%
\belowdisplayshortskip 7pt plus3.5pt minus0pt
\def\@listi{\parsep 4.5pt plus 2pt minus 1pt
      \itemsep \parsep
      \topsep 9pt plus 3pt minus 3pt}}

\relax

\def\section{\@startsection{section}{1}{\z@}{3.5ex plus 1ex minus  .2ex}{2.3ex plus .2ex}{\large\bf}}

\def\thesection{\arabic{section}}
\def\thesubsection{\arabic{section}.\arabic{subsection}}

\def\appendix{\setcounter{section}{0}
 \def\thesection{Appendix \Alph{section}}
 \def\thesubsection{\Alph{section}.\arabic{subsection}}
 \def\theequation{\Alph{section}.\arabic{equation}}}
\renewcommand{\theequation}{\arabic{section}.\arabic{equation}}

}{
\renewcommand{\theequation}{\arabic{equation}}

} %se letter=0, commenta questo

\begin{document}
% ========================================================================
\def\thefootnote{\fnsymbol{footnote}}
\def\thetitle{Spectral Walls at One Loop }
\def\autone{Jarah Evslin}
\def\auttwo{Chris Halcrow}
\def\autthree{Tomasz Roma\'nczukiewicz}
\def\autfour{Andrzej Wereszczy\'nski}
\def\affa{Institute of Modern Physics, NanChangLu 509, Lanzhou 730000, China}
\def\affb{University of the Chinese Academy of Sciences, YuQuanLu 19A, Beijing 100049, China}
\def\affc{School of Mathematics, University of Leeds, Leeds LS2 9JT, United Kingdom}
\def\affd{Institute  of  Physics,  Jagiellonian  University,  Lojasiewicza  11,  Krak\'ow,  Poland}

\ifthenelse{\equal{\pr}{1}}{
\title{\thetitle}
\author{\autone}
\author{\auttwo}
\affiliation {\affa}
\affiliation {\affb}
\affiliation {\affc}
%pr e uno

}{

\begin{center}
{\large {\bf \thetitle}}

\bigskip

\bigskip

%\catcode`@=11

{\large \noindent  \autone{${}^{1,2}$}, \auttwo{${}^{3}$}, \autthree{${}^{4}$} and \autfour{${}^{4}$}}

%{\large \noindent  \autone{${}^{1,2}$} \footnote{jarah@impcas.ac.cn} and \auttwo{${}^{1,2}$} \footnote{guohengyuan@impcas.ac.cn}}

\vskip.7cm

1) \affa\\
2) \affb\\
3) \affc\\
4) \affd\\

\end{center}

}

\begin{abstract}
\noindent
A spectral wall is a surface in a moduli space of classically BPS solitons where an internal excitation crosses the continuum mass threshold.  It has recently been shown that spectral walls in classical field theory repel solitons whose corresponding internal excitations are excited.  We investigate, for the first time, the fate of spectral walls in a quantum theory, calculating the instantaneous acceleration of a wave packet of classically BPS antikinks in the presence of an impurity.  We find that, in the quantum theory, the antikinks are repelled even when the bound mode is not excited. Perhaps due to this repulsive force, the dramatic classical effects of the spectral wall are not seen in our quantum calculation. 
\end{abstract}

% \vfill
%
% \end{titlepage}
\setcounter{footnote}{0}
\renewcommand{\thefootnote}{\arabic{footnote}}

\ifthenelse{\equal{\pr}{1}}{
\maketitle
}{}

%%%%%%%%%%%%%%%%%%%%%%%%%%%%%%%%%%
\section{Introduction}
%%%%%%%%%%%%%%%%%%%%%%%%%%%%%%%%%%

Topological solitons are stable, localized, particle-like solutions of nonlinear partial differential equations which carry a quantized amount of the pertinent topological charge. In $(1+1)D$ the best studied solitons are kinks. The first modern study of kinks as solitons was done  in the sine-Gordon model by Skyrme and Perring, as a toy model of the eponymous $(3+1)D$ Skyrme model \cite{SkPerr}. Since then, kinks have served as a simple tool to study difficult problems in solitonic systems. Even in these simple kink systems, highly complicated classical dynamics has been seen \cite{MS,Sh,K}. For instance, kinks and antikinks can either annihilate or separate after colliding. These two possibilities form a fascinating fractal pattern depending on the parameters (e.g., velocity) of the initially colliding solitons \cite{Sug, Mosh, CSW}. This phenomena has only recently been understood in an effective collective coordinate approximation \cite{MORW}.

Recently, an intriguing phenomenon called a {\it spectral wall} has been discovered \cite{muri}. It affects the dynamics of solitons in BPS theories, where classically there is no force between static solitons (or between a soliton and an impurity). In these rather rare and important models, topological solitons can be placed at any distance from each other \cite{Bo, JT, Ma}. This leads to a nontrivial moduli space of the energetically equivalent solutions whose lowest order dynamics can be  accurately described in terms of geodesic flow on the canonical moduli space \cite{NM, AH}. However, even though the classical solutions possess the same energy, the spectrum of their small perturbations can differ. That is to say that, depending on the mutual distance between the kink and impurity, the structure of the normal modes changes. At a given distance one of the massive bound modes can reach the mass threshold, at which the continuous spectrum begins, where it transmutes into a non-normalizable threshold mode \cite{muri}. We call this point the spectral wall. Importantly, this mode transmutation acts as an obstacle in the solitonic dynamics \cite{muri}. At the spectral wall the soliton can: form a stationary state, be reflected or pass through the spectral wall with a temporal distortion. The outcome depends on the value of the amplitude of the mode which enters the continuum. In any case, the spectral wall seems to play a critical role in the dynamics of excited BPS solitons. This new dynamical phenomena should be a generic feature of soliton dynamics but was first discovered using kinks.

Kink systems are also a fertile place to study quantum corrections to solitons, which have proven immensely difficult to calculate in more than one dimension.  Over the years, many techniques have been introduced to calculate one-loop corrections to kinks, beginning with the semiclassical approach of Ref.~\cite{dhn2}.  At one loop, kinks are described by a free theory and so the dynamics becomes more rich at two loops \cite{crystal}.  The multiloop dynamics is usually treated using the collective coordinate approach of Ref.~\cite{gjs}.  This approach is very powerful, but requires a nonlinear canonical transformation which is so complicated that it has hindered progress in the field.  

In Ref.~\cite{me2loop}, a much more economical approach to multiloop calculations has been introduced, building upon the manifestly finite, Hamiltonian, one-loop approach of Refs.~\cite{cahill76,mekink}.  In this approach, the quantum field is expanded about a classical solution at a fixed base point in moduli space.  Besides being manifestly UV finite, the main advantage is that the treatment is fully linear, allowing access to problems \cite{mephi42,meunbind} which would be prohibitively difficult with traditional methods.  The limitation is that it can only treat kinks near a base point, which is arbitrary but fixed in each calculation.  Thus instantaneous accelerations can be computed, but following the motion of a kink over a macroscopic distance would require a gluing of results obtained at distinct base points.

In this paper we use this new approach to loop calculations to investigate the fate of spectral walls when one-loop quantum corrections are taken into account. Looking from a wider perspective we also want to understand the consequences, if any, of a transition of a normal mode to the continuum spectrum in the quantum version of a BPS theory.   We use the formalism of Refs.~\cite{me2loop, mekink}.  After introducing our model in Sec.~\ref{sec:imp}, the generalization of this formalism to theories with impurities is introduced in Sec.~\ref{unloopsez}, where it is applied to compute one-loop corrections to kink energies.  Unlike previous applications of this formalism, due to the lack of translation invariance in this model, it is essential that kinks are treated properly as wave packets, as is described in Sec.~\ref{smearsez}.  Finally these results are assembled in Sec.~\ref{dynsez} to describe the quantum dynamics of kinks in these theories, culminating in the calculation of the instantaneous acceleration caused by the impurity.

%%%%%%%%%%%%%%%%%%%%%%%%%%%%%%%%%%
\section{The BPS Kink-Impurity Model} \label{sec:imp}
%%%%%%%%%%%%%%%%%%%%%%%%%%%%%%%%%%
In this section we will introduce the simplest field theoretical model which admits spectral walls \cite{muri}. However, we underline that this phenomenon occurs in a large variety of field theories including ones with no impurity.
%%%%%%%%%%%%%%%%%%%%%%%%%%%%%%%%%%
\subsection{The BPS solutions}
%%%%%%%%%%%%%%%%%%%%%%%%%%%%%%%%%%
Let us begin with the standard scalar field model in (1+1) dimensions with at least a double vacuum potential $V(\phi)$
\beq \label{eq:Lag}
L=\int_{-\infty}^\infty \left( \frac{1}{2} (\partial_\mu \phi)^2 - V(\phi)\right) dx.
\eeq
The BPS sector of the classical theory consists of field configurations which satisfy one of the two static Bogomolny equations
\beq \label{eq:bogo}
\frac{d\phi}{dx} = \pm \sqrt{2V}.
\eeq
Generally, these equations admit solutions with topological charge $\pm 1$ called (anti)kinks which we denote $\Phi(x)$. These are the lowest energy field configurations with topological charge $\pm 1$. By studying the field structure or the energy density of the (anti)kink, we can associate a position $x_0\in\mathbb{R}$ to any solution $\Phi_{x_0}(x)$.  In the numerical calculations below we define the $x_0$ label of each solution to be the unique zero of that solution
\beq
\Phi_{x_0}(x_0)=0
\eeq
and we will always choose parameters such that there is only one zero.  Due to the translation invariance of the Lagrangian \eqref{eq:Lag} we can shift the position of the (anti)kink by applying the transformation
\beq
\Phi_{x_0}(x)\rightarrow\Phi_{x_0+c}(x)=\Phi_{x_0}(x-c). \label{tsim}
\eeq
Hence, there is a one dimensional moduli space of energetically equivalent BPS solutions. There are also trivial solutions to \eqref{eq:bogo}, the topologically trivial vacua.

Typically the addition of an impurity $\sigma(x)$, {\it{i.e.}} a background, non-dynamical field, introduces an interaction in the single soliton sector. 
This means that there is a static force between the (anti)kink and impurity. As a consequence, there is a preferred position of the soliton with respect to the impurity where the energy takes its minimal value. Hence, the symmetry (\ref{tsim}) is broken and the BPS sector trivializes to only one (or perhaps a finite number of) solutions. This resembles the case of a kink-antikink pair, which is typically not a member of any BPS sector as it does not solve any static Bogomolny equation.  As a result, kinks and antikinks feel a mutual static force. 

It has recently been demonstrated that it is possible to couple a scalar field to an impurity in a BPS-preserving manner \cite{muri, impurita}. This works for Lagrangians of the form
\beq
L=\int_{-\infty}^\infty \left( \frac{1}{2} (\partial_\mu \phi)^2 - V(\phi, x) + \sqrt{2} \phi \partial_x \sigma(x)\right) dx,
\eeq \label{eq:laggenv}
where
\beq
V(\phi, x)= \left( W(\phi) +\sigma(x)\right)^2
\eeq
is the impurity-deformed potential. For other possibilities see \cite{susy, solvable, azadeh, susy-2}. Then, although (\ref{tsim}) remains broken, there is nonetheless one Bogomolny equation
\beq
\frac{d\phi}{dx} = - \sqrt{2}(W(\phi) + \sigma(x)), \label{bps-kin}
\eeq
which admits infinitely many, energetically equivalent solitonic (here, antikink) solutions, $\Phi_{x_0}(x)$, parametriz\-ed again by the continuous real parameter $x_0$, which can be interpreted as the position of the antikink. Since all solutions have the same energy, there is no static force between the antikink and the impurity. The two can be located any distance from each other. It is straightforward to show that solutions of the Bogomolny equation obey the full Euler-Lagrange equation of motion arising from (\ref{eq:laggenv}).

For an explicit example, consider the Lagrangian
\beq
L=\int_{-\infty}^\infty \left( \frac{1}{2} (\partial_\mu \phi)^2 - \frac{m^2}{4\lambda} (W(\phi) + \sigma)^2 +  \frac{m}{ \sqrt{2\lambda}} \phi \partial_x \sigma(x)\right) dx, \label{Lag-imp}
\eeq
with 
%\beq \label{imp-exm}
%W(\phi)=\frac{m \left(1-\lambda \phi^2 \right)}{2\sqrt{2\lambda}}\hsp
%\sigma(x)=\frac{ m \alpha}{\sqrt{2\lambda}}  \sech^2(Mx)
%\eeq
\beq \label{imp-exm}
W(\phi)=\frac{ \left(1-\lambda \phi^2 \right)}{\sqrt{2}}\hsp
\sigma(x)= \alpha  \sech^2(lx)
\eeq
which is a BPS-impurity deformation of the $\phi^4$ model. 

Note that $\lambda^{-1}$ has dimensions of action, and so $\hbar\lambda$ is the dimensionless constant. A power series expansion in this parameter can be interpreted as a semiclassical expansion in $\hbar$ or a perturbative expansion in $\lambda$.  In the presence of an impurity there are other dimensionless quantities, such as $m/l$ and $\alpha$ which will play important roles below.  %Setting $\hbar=1$ we conclude that the semiclassical expansion is equivalent to a double expansion in $\lambda/m^2$ and in $m^2/\alpha^2$.   In particular, for $l\sim m$, it is reliable if $\lambda<<m^2<<\alpha^2$.   We leave an analysis of the $m/\alpha$ expansion and its implications to future work.%This condition is satisfied for the $\alpha=3$ case considered below, and it would be interesting to investigate higher order corrections to the $\alpha=0.3$ case.

\begin{figure} [h]%[!tph]
	\begin{center}
\includegraphics[width=3.2in]{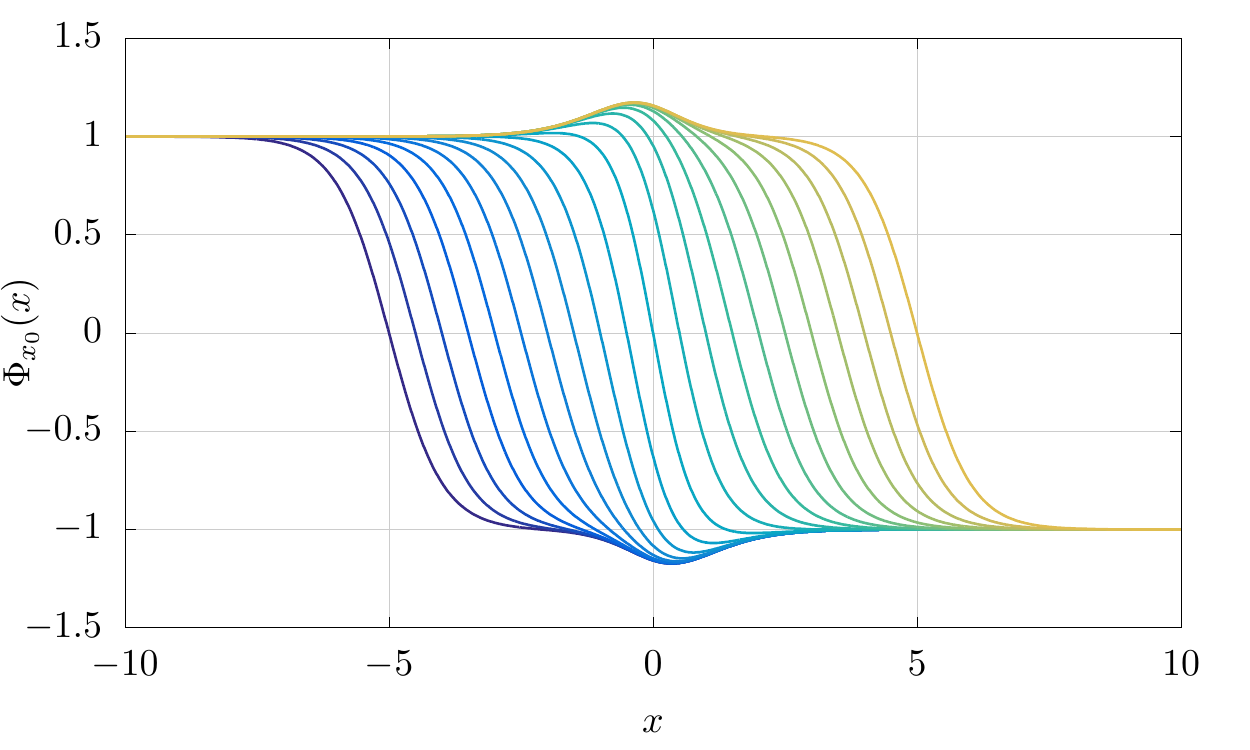}
\includegraphics[width=3.2in]{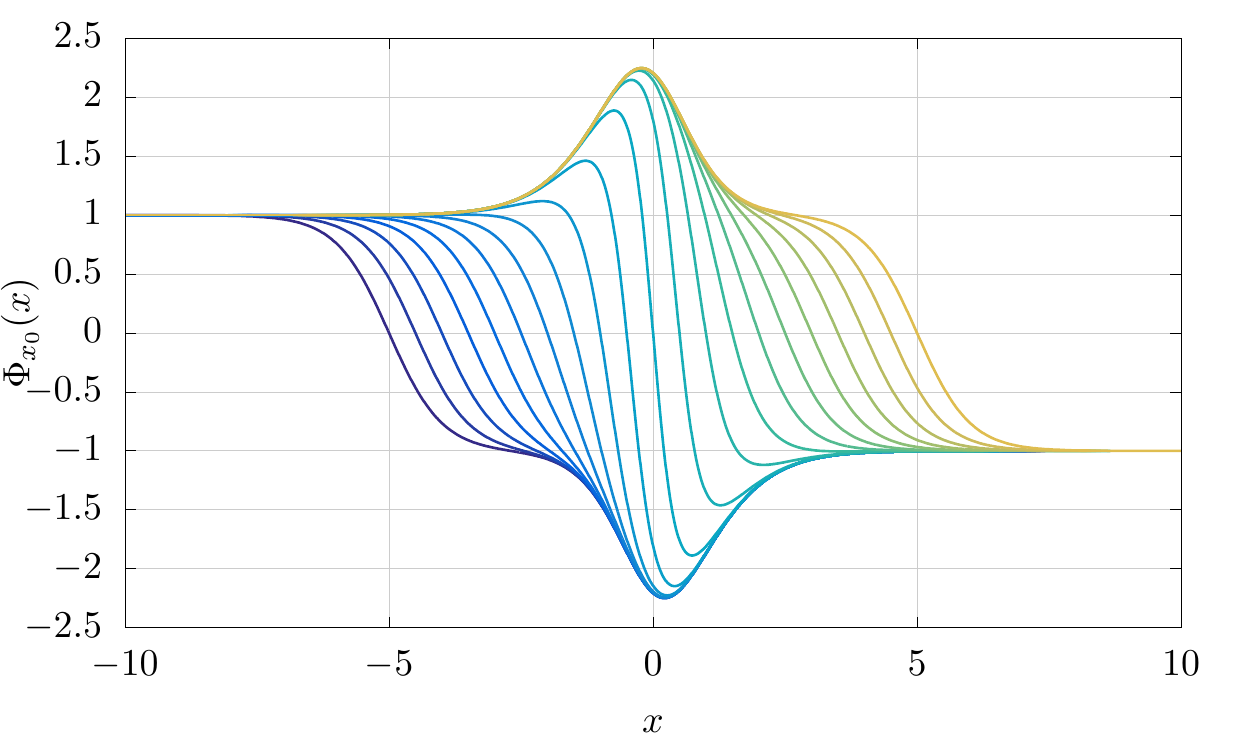}
		\caption{Antikinks $\Phi_{x_0}(x)$  at $\alpha$ equal to $0.3$ (left) and $3.0$ (right) for several values of $x_0$.}
		\label{fig:kinks}
	\end{center}
\end{figure}

Specifically, in our numerical computations the parameters take the values $m=2$, $\lambda=l=1$ and $\alpha=0.3$ or $\alpha=3.0$. Then, the BPS antikink solution has the classical energy $Q_0=4/3$. We plot the BPS solutions in Fig. \ref{fig:kinks}. Note that, although the basic translation symmetry (\ref{tsim}) is broken, we still label the solutions by $x_0$ and interpret this as the position of the antikink. When the antikink center is far from the impurity, the configuration looks like an antikink-without-impurity superposed with the vacuum solution near the impurity. Near the impurity, the antikink is deformed. 

The one-parameter family of energetically equivalent antikink-impurity solutions $\Phi_{x_0}(x)$ forms the canonical moduli space, which describes the simplest dynamics of an antikink passing through the impurity. Concretely, the static BPS solutions $\Phi_{x_0}(x)$ are inserted into the original Lagrangian with the modulus $x_0$ promoted to a time dependent variable. Then, we arrive at the following collective coordinate model 
\beq
L(x_0)=\frac{1}{2} M(x_0) \dot{x}_0^2,
\eeq
where the dot denotes the time derivative and $M(x_0)$ is the metric on a 1-dimensional moduli space
\beq
M(x_0)=\int_{-\infty}^\infty \left( \partial_{x_0} \Phi_{x_0}(x)\right)^2 dx.
\eeq
This function can also be interpreted as a generalized mass of the soliton, which changes in the vicinity of the impurity. Asymptotically, as $x_0 \to \pm \infty$, it tends to the mass of the soliton in the $\phi^4$ model
\beq
M_0=\frac{m}{\sqrt{\lambda}} \int_{-1}^1 d\phi W(\phi) = \frac{2}{3} \frac{m}{\lambda}
\eeq
which, for our choice of parameters, is $M_0=4/3$ and equals the energy of the soliton $Q_0=M_0$. More generally, due to the presence of the impurity, $M(x_0)\neq Q_0$. In this approximation, $x_0$ is the kinetic degree of freedom of the BPS antikink. The resulting time dependence of $x_0$ models the collision of the antikink with the impurity very well, provided no internal modes are excited and the initial velocity is small, $v<<1$ \cite{impurita}. 

The Bogomolny equation also supports two topologically trivial solutions, i.e., vacua with vanishing classical energy $Q_0=0$. For the assumed impurity they are isolated solutions, which do not span any nontrivial moduli space. Formally, the vacuum moduli space is just a point in each vacuum sector. 
%%%%%%%%%%%%%%%%%%%%%%%%%%%%%%%%%%
\subsection{The Mode Structure}
%%%%%%%%%%%%%%%%%%%%%%%%%%%%%%%%%%
A soliton can also store energy in internal degrees of freedom, which are typically identified with the massive normal modes that arise in the linear perturbation theory. Concretely, we perturb the static BPS solutions, $\Phi(x)$, which include both antikink and topologically trivial states, by a small perturbation 
\beq
\phi(x,t)=\Phi_{x_0}(x)+e^{-i\omega t}g(x) 
\eeq
which, when inserted into the equation of motion, leads to the small mode equation 
\beq
0=\left(\omega^2-\v2\right)g(x)+\partial_x^2 g(x) \label{cleq}
\eeq
where we have defined
\beq
\v{n}=\frac{\delta^n}{(\delta\phi(x))^n}V(\phi(x),x)|_{\phi(x)=\Phi_{}(x)}.
\eeq
Assuming that $\omega \in \mathbb{R}$ and imposing a normalization condition on the mode (wave) function $g$ leads to three possibilities. First of all, there can be a solution with zero frequency referred to as the {\it zero mode}.  This solution exists when there is a continuous family of energetically equivalent solutions, for example the BPS antikink solutions $\Phi_{x_0}(x)$. Hence, the zero mode is intimately related to a nontrivial moduli space. Furthermore, its excitation corresponds to a kinetic motion of the soliton. In our case, the zero mode is just
\beq
g_B=\frac{1}{\sqrt{M(x_0)}} \partial_{x_0} \Phi_{x_0}(x),
\eeq
which obeys the normalization condition
\beq
\int dx g_B^2(x)=1. \label{gnorm}
\eeq
The vacuum solutions do not support any zero modes as the moduli space is discrete.

In addition there may be a set of discrete massive (bound) modes, which we call {\it shape modes} $g_S(x)$ with $0<\omega_S<m$, where $m^2$ is the {\it mass threshold} defined as the asymptotic value of $\v2$ at $x = \pm \infty$ \footnote{ If these asymptotic values are not equal, the one-loop vacuum energies will differ on the two sides of the kink \cite{wpol}.  As a result the kink, being a wall separating the true and false vacua, will accelerate \cite{tstabile,wstabile} and so its mass is difficult to define. }.  Here $g_S(x)$ is real and fulfils the same normalization as the zero mode (\ref{gnorm}). 

Finally, there are also {\it continuum} solutions at each $\omega\geq m$, describing {\it radiation}.  For each $\omega>m$, let $k=\sqrt{\omega_{k}^2-m^2}>0$. There are two real, linearly independent solutions at each frequency $\omega_k$.  At large $|x|$ or $|k|$ these tend to sine-waves of wave number $k$.  Let the $\psi_1(x)$ and $\psi_2(x)$ be an orthonormal basis of the space of real frequency $\omega_k$ solutions, chosen to have amplitude 1 at $x\rightarrow\pm\infty$.  Then define
\beq
g_{\pm k}(x)=\psi_1(x)\pm i \psi_2(x).
\eeq
These will also be orthonormal and satisfy the normalization condition
\beq
\int dx g^*_{k_1}(x) g_{k_2}(x)=2\pi\delta(k_1-k_2).
\eeq

The space of linear perturbations  is even richer if the normalization condition or the $\omega \in \mathbb{R}$ condition is relaxed. There is a non-normalizable {\it threshold mode} with $\omega=m$, whose wave function approaches a constant asymptotically. There may also be {\it antibound modes}, which are non-normalizable modes with purely real frequency and an exponentially growing wave function at spatial infinity. Finally, there can be genuine {\it quasi-normal modes} (resonances) which are solutions of the linear mode equation with complex frequencies $\Omega = \omega + i\Gamma$ and purely out-going boundary conditions
\beq
\stackrel{\rm{lim}}{{}_{x\rightarrow -\infty}}\partial_x{\rm{Arg}}(g(x))>0>\stackrel{\rm{lim}}{{}_{x\rightarrow +\infty}}\partial_x{\rm{Arg}}(g(x))
.
\eeq
Physically, $\Gamma$ encodes information about the decay of this mode. All these modes may actively participate in the dynamics of kinks. 

Equation (\ref{cleq}) is a special case of the Sturm-Liouville equation.  Therefore, on any finite interval, the normal modes are a basis of the space of functions.  We are interested in the entire real line, not a finite interval.  Here the space of normal modes $g_B(x)$, $g_S(x)$ and $g_k(x)$ is a basis of all $\delta$-function normalizable functions. Quasi-normal modes are not in this space.  The corresponding completeness relation is
\beq
g_B(x)g_B(y)+  g_S(x)g_S(y)+ \int dk g_k^*(x)g_k(y)=\delta(x-y). \label{crel}
\eeq
Note that if there are multiple shape modes $S_i$, one should simply sum over them $\sum_i  g_{S_i}(x)g_{S_i}(y)$.  Similarly, if there are no zero modes or shape modes, the corresponding term should be omitted from the sum.

\begin{figure}% [h]%[!tph]
	\begin{center}
\includegraphics[width=3.2in]{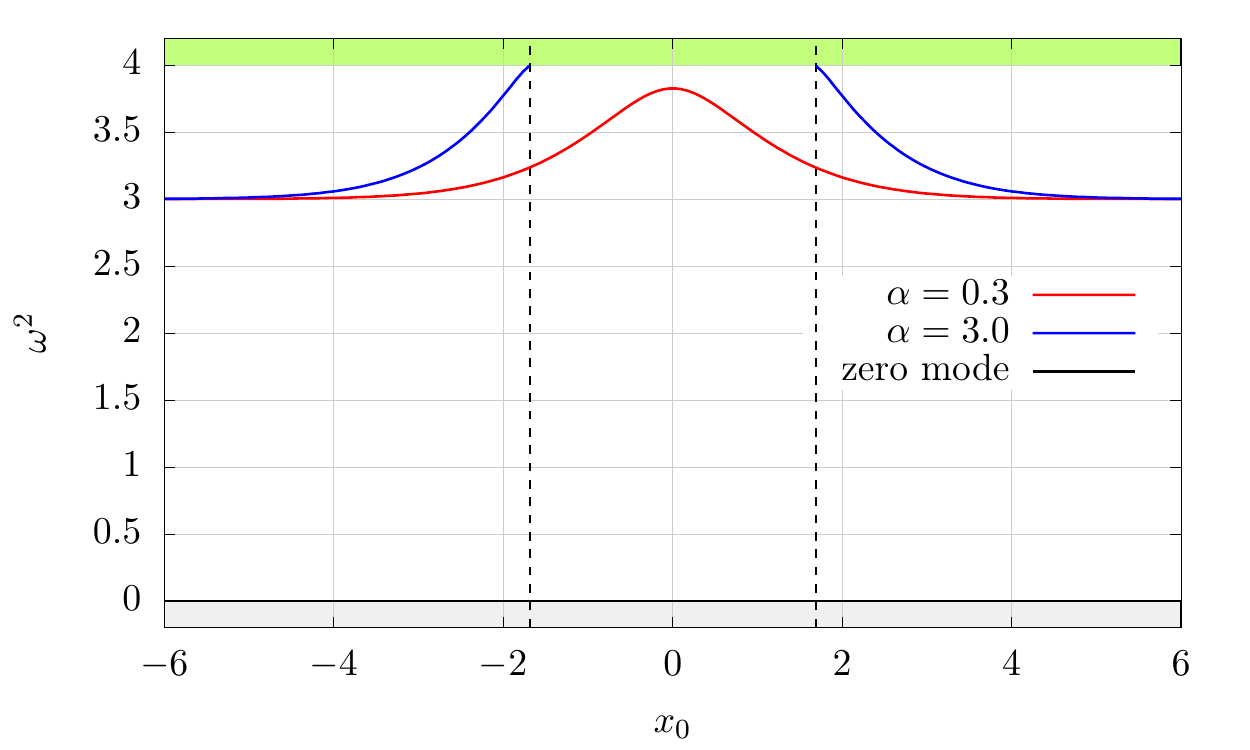}
	\includegraphics[width=3.0in]{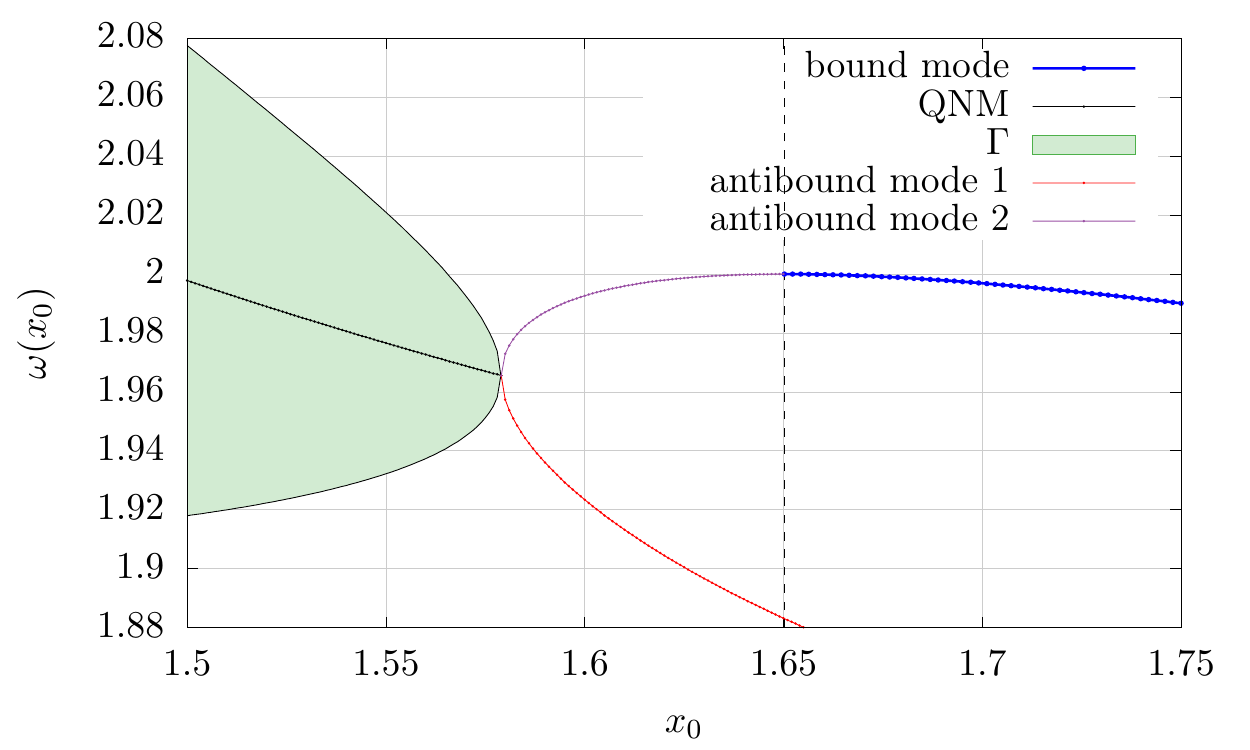}
		\caption{The structure of normal modes in the BPS impurity kink model (\ref{Lag-imp})-(\ref{imp-exm}) for the BPS antikink solution $\Phi_{x_0}(x)$. {\it Left}: Change of the frequency of the normal modes for two values of impurity strength as a function of modulus $x_0$. There is always a zero mode, while the shape mode may reach the mass threshold, $\omega^2=4$, above which the continuum spectrum of scattering modes (radiation) begins. {\it Right}: The linear modes in the vicinity of the spectral wall for $\alpha=3.0$. }
		\label{omfig}
	\end{center}
\end{figure}

An important feature of the BPS-impurity model is that, although the BPS solutions $\Phi_{x_0}(x)$ are energetically equivalent, they do not have the same spectrum of linear perturbations. Indeed, the structure of the linear modes depends on the parameter $x_0$, which characterizes the soliton-impurity distance. This property occurs in more complicated BPS systems, e.g., multi-scalar BPS models and vortices in the Abelian Higgs model at critical coupling. However, the BPS-impurity systems are the simplest field theoretical set-ups with a flow of the spectral structure. In Fig. \ref{omfig}, left panel, we plot the normal mode structure for our example (\ref{Lag-imp})-(\ref{imp-exm}). 
If the impurity is small, $\alpha = 0.3$, the frequency of the unique shape mode is only moderately changed as the antikink approaches the impurity. On the other hand, for a stronger impurity, $\alpha=3$, there is a dramatic change in the spectral structure at $|x_0|=x_\text{sw} = 1.650231$. When $|x_0| < x_\text{sw}$ there is no shape normal mode in the spectrum. The point $x_\text{sw}$ defines the position of the {\it spectral wall}, a phenomenon which we explain below.

A detailed study reveals a rather involved mode structure as we pass through the spectral wall, shown in Fig. \ref{omfig}, right panel. We see the following behavior: at the spectral wall the shape mode becomes a non-normalizable threshold mode with frequency $\omega = 2$. As $|x_0|$ decreases, it transmutes into an antibound mode. There is another antibound mode with lower frequency, which also exists in the region outside the spectral wall $|x_0| > x_{\text{sw}}$. As $x_0$ further decreases the antibound modes meet and combine into a quasi-normal mode. Now, both $\omega$ and $\Gamma$ grow as $|x_0|$ decreases. Our detailed analysis goes beyond the initial study of this model in \cite{muri}. 

%%%%%%%%%%%%%%%%%%%%%%%%%%%%%%%%%%
\subsection{The Spectral Wall}
%%%%%%%%%%%%%%%%%%%%%%%%%%%%%%%%%%

\begin{figure}% [h]%[!tph]
		\begin{center}
\includegraphics[width=3.2in]{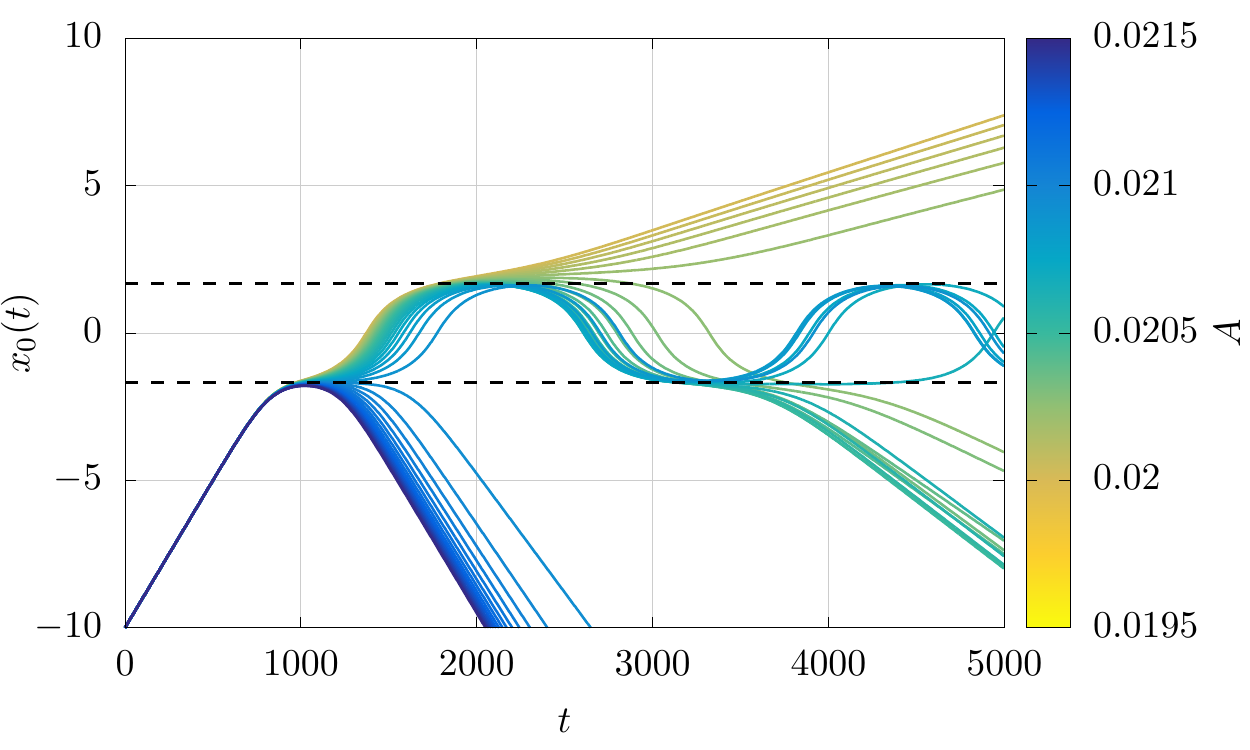}
\includegraphics[width=3.2in]{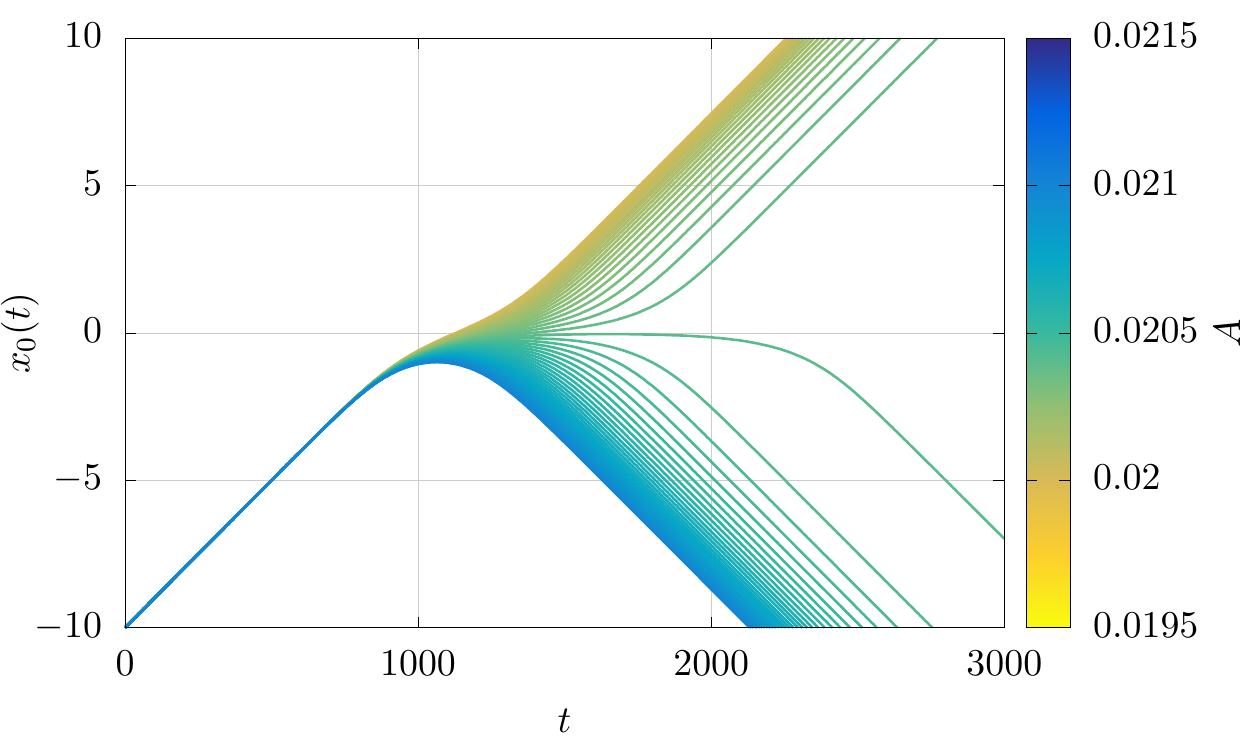}
		\caption{Classical dynamical evolution of $x_0$, the position of the antikink, with an impurity near the critical normal mode excitation $A_{crit}$ for $\alpha=3$ (left) and $\alpha=0.3$ (right). A spectral wall only exists for the $\alpha=3$ system. The dynamics is heavily influenced by the spectral wall at $x_{sw}=1.65$ (dotted line, left). }
		\label{fig:sw}
	\end{center}
\end{figure}

The fact that a shape mode reaches the mass threshold is not only a mathematical peculiarity of the model but has a profound effect on the dynamics of the BPS antikink, known as the spectral wall phenomenon \cite{muri}. The transition of the shape mode %into a non-normalizable threshold mode 
creates an obstacle to an excited soliton's motion at $x_\text{sw}$. This is seen in Fig. \ref{fig:sw} where the dynamics of an excited antikink for $\alpha=3$ (when a spectral wall exists, left panel), and $\alpha=0.3$ (when there's no spectral wall, right panel) is presented. Strictly speaking, if the amplitude of the shape mode is excited at the critical amplitude, $A=A_{crit}$ then the antikink forms a stationary, arbitrary long lived state at $x_0=x_{sw}$. If $A<A_{crit}$, it passes the spectral wall and its motion is less and less affected as the amplitude decreases. If $A>A_{crit}$, the soliton is reflected and the reflection point occurs sooner, i.e., at bigger distance, as the amplitude grows. This behavior was also observed in a BPS two-scalar field model \cite{sw-2fields} and is expected in any BPS system provided there is a mode crossing the mass threshold, e.g., \cite{eg-sw-1}-\cite{eg-sw-6}. 

The spectral wall can be studied in an extended version of the moduli space approximation where, together with the BPS solutions, shape modes are included. This is sometimes referred as a vibrational moduli space and it describes the motion of a soliton where both kinetic (zero modes) as well as internal (shape modes) degrees of freedom are taken into account. The resulting vibrational moduli space has an essential singularity at the location of the spectral wall, reflecting the transmutation of the shape mode into a non-normalizable threshold mode. 

In the subsequent sections we will study one-loop quantum corrections to the BPS-impurity model, hoping to understand the quantum spectral wall phenomenon. As the transition of a normal mode to the continuous spectrum has a rather big impact on classical soliton dynamics, one may hope to see such an impact on the quantum dynamics too.

%\newpage
%%%%%%%%%%%%%%%%%%%%%%%%%%%%%%%%%%
\section{One-Loop Corrections to the BPS Antikink-Impurity Bound State} \label{unloopsez}
%%%%%%%%%%%%%%%%%%%%%%%%%%%%%%%%%%

To calculate the one-loop correction to the antikink, we will need to generalize the formalism developed in Ref.~\cite{me2loop}. This is because our system does not have simple translation invariance and the small-fluctuation potential is not symmetric in $x$. We will deal with these technical issues in Subsects. \ref{sec:kinkham}-\ref{sec:oneloop} before applying the formalism to the antikink-impurity model in Subsec. \ref{kinksez}.

%%%%%%%%%%%%%%%%%%%%%%%%%%%%%%%%%%
\subsection{The Kink Hamiltonian} \label{sec:kinkham}
%%%%%%%%%%%%%%%%%%%%%%%%%%%%%%%%%%
Let us start with the Hamiltonian of the BPS-impurity model (\ref{eq:laggenv}), with an arbitrary (two vacuum) $W$ and impurity $\sigma(x)$
\beq
H=\int dx :\ch(x):_a\hsp
\ch(x)=\frac{1}{2}\left[\pi^2(x)+\left(\partial_x\phi(x)\right)^2\right]+V(\phi,x)-\sqrt{2}\phi(x)\partial_x\sigma(x)
\eeq
where $\pi(x)$ is the conjugate momentum, and the plane-wave normal-ordering $::_a$ will be defined in Subsec.~\ref{contsez}. %We will use units where $\lambda=1$, but reinsert this dimensionless constant when we do the semiclassical expansion in Sec. \ref{sec:doingexpansion}.
Choose any time-independent classical solution $\phi(x)=\Phi(x)$ of the static equation of motion
\beq
\partial_x^2 \Phi_{}(x)=\v1-\sqrt{2}\partial_x\sigma(x). \label{feq}
\eeq
Let us define the unitary displacement operator
\beq
\mathcal{D}_\Phi={\rm{exp}}\left(-i\int dx \Phi(x)\pi(x)\right) \label{df}
\eeq
and the similarity-transformed Hamiltonian
\beq
H\p=\mathcal{D}_\Phi^\dag H\mathcal{D}_\Phi.
 \label{hpd}
\eeq
We will see that perturbation theory using $H\p$ can be used to build the Fock space of perturbative excitations above the ground state of the sector of the quantum field theory corresponding to the classical solution $\Phi(x)$.  We refer to the space of such states as the $\Phi(x)$-sector.  Following the usual convention, we will refer to $H\p$ as the {\it{kink Hamiltonian}} because we will usually be interested in the case in which $\Phi(x)$ is a BPS antikink solution $\Phi_{x_0}(x)$, and so $H\p$ will be used to construct antikink sector states.   However in \ref{vacsez} we will apply this formalism to the case when $\Phi(x)$ is a topologically trivial solution, in which case $H\p$ can be used to construct vacuum sector states.

The similarity transformation (\ref{hpd}) commutes with any normal ordering prescription \cite{mekink}.  As $H$ and $H\p$ are similar, they have the same spectra.  Therefore for every eigenvector $|\Psi\rangle$ of $H$ with energy $E$, $\mathcal{D}_\Phi^\dag|\Psi\rangle$ is an eigenvector of $H\p$ with the same energy
\beq
H\p\mathcal{D}_\Phi^\dag|\Psi\rangle=E\mathcal{D}_\Phi^\dag|\Psi\rangle.
\eeq
In other words, the energy $E$ is an eigenvalue of both $H\p$ or $H$, and so it may be obtained by solving the eigenvalue problem for either operator.  The reason that one introduces a kink Hamiltonian $H\p$ is that its eigenvectors corresponding to $\Phi(x)$-sector states, and their eigenvalues, can be obtained in perturbation theory. 

This situation can be summarized as follows.  The theory is defined by a Hamiltonian $H$ and a Hilbert space of states.  Finding (anti)kink states directly in this space would be a difficult, nonperturbative problem.  Thus one first transforms the Hilbert space using the unitary operator $\mathcal{D}_\Phi^\dag$ where $\Phi(x)$ is the classical solution corresponding to the sector of interest.  Intuitively, this removes the classical part from the field $\phi(x)$, leaving only the quantum fluctuations $\phi(x)-\Phi(x)$.  Then one diagonalizes $H\p$ perturbatively, which yields the Fock space of $\Phi$-sector states.  Finally, to obtain the corresponding states in the original Hilbert space, one acts with $\mathcal{D}_\Phi$.

Now, we expand
\beq
H\p=\sum_{j=0} H\p_j\hsp
H\p_j=\int dx \ch_j(x)
\eeq
where $\ch_j(x)$ contains all terms of order $j$ in the fluctuations when normal-ordered using $::_a$. 
The first term $H\p_0$ is just the classical energy $Q_0$ of the solution $\Phi(x)$.  
The tree-level tadpole $H\p_1$ vanishes by the classical equations of motion
\beq
\ch_1(x)=\left[-\partial_x^2 \Phi(x)+\v1-\sqrt{2}\partial_x\sigma(x)\right]\phi(x)=0.
\eeq
The next contribution yields the one-loop quantum correction to the energy of the $\Phi(x)$-sector ground state. It is determined by the free Hamiltonian $H_2\p$, according to the dimensional analysis of Ref.~\cite{dhn2}.  The free Hamiltonian density is
\beq \label{eq:H2den}
\ch_2(x)=\frac{1}{2}:\left[
\pi^2(x)+\left(\partial_x\phi(x)\right)^2+\v2 \phi^2(x)
\right]:_a, 
\eeq
which leads to the equation (\ref{cleq}) satisfied by small fluctuations (linear modes). 

%%%%%%%%%%%%%%%%%%%%%%%%%%%%%%%%%%
\subsection{From Plane-Wave to Normal-Mode Normal-Ordering} \label{contsez}
%%%%%%%%%%%%%%%%%%%%%%%%%%%%%%%%%%

We will work in the Schr\"odinger picture, so that the fields $\phi(x)$ and $\pi(x)$ are independent of time and satisfy the canonical commutation relations
\beq
[\phi(x),\pi(y)]=i\delta(x-y). \label{ccr}
\eeq
Together the operators $\phi(x)$ and $\pi(x)$ generate the operator algebra of a scalar theory.  We will need several different bases of this algebra.  A decomposition in terms of plane waves yields the basis
\beq
\phi_p=\int dx \phi(x)e^{ipx}\hsp
\pi_p=\int dx \pi(x)e^{ipx}
\eeq
while a decomposition in normal modes yields
\beq
\phi_k=\int dx \phi(x) g^*_k(x)\hsp
\pi_k=\int dx \pi(x) g^*_k(x)
\eeq
where $k$ runs over all real numbers representing continuum states as well as the discrete values $S$ representing shape modes.  For the zero-mode we will use the standard notation
\beq
\phi_0=\int dx \phi(x) g_B(x)\hsp
\pi_0=\int dx \pi(x) g_B(x).
\eeq
These bases can be distinguished because we will always use $p$ and $q$ to label plane waves and $k$ to label normal modes.  The commutation relations (\ref{ccr}) imply the relations
\beq
[\phi_p,\pi_q]=2\pi i \delta(p+q)\hsp
[\phi_{k_1},\pi_{k_2}]=2\pi i \delta(k_1+k_2)\hsp [\phi_I,\pi_J]=i\delta_{IJ}
\eeq
where $I$ and $J$ run over bound states $B$ and $S$.  The decompositions may be inverted using the completeness of the plane waves and the normal modes (\ref{crel})
\bea
\phi(x)&=&\pin{p}\phi_p e^{-ipx}=\phi_0g_B(x)+\phi_Sg_S(x)+\pin{k}\phi_k g_k(x)\\
\pi(x)&=&\pin{p}\pi_p e^{-ipx}=\pi_0g_B(x)+\pi_Sg_S(x)+\pin{k}\pi_k g_k(x)\nonumber
\eea
where for concreteness we have considered a single shape mode.

These bases can be rearranged as usual to construct creation and annihilation operators.  In the case of the plane wave basis
\beq
A^\dag_p=\frac{\phi_p}{2}-i\frac{\pi_p}{2\omega_p}\hsp
\frac{A_{-p}}{2\omega_p}=\frac{\phi_p}{2}+i\frac{\pi_p}{2\omega_p}.
\eeq
Note that $A^\dag$ and $A$, so normalized, are only Hermitian conjugate after a rescaling.  One can easily check that these satisfy the Heisenberg algebra
\beq
[A_p,A^\dag_q]=2\pi \delta(p-q).
\eeq
Similarly the normal mode basis can be used to construct
\beq
B^\dag_k=\frac{\phi_k}{2}-i\frac{\pi_k}{2\omega_k}\hsp
\frac{B_{-k}}{2\omega_k}=\frac{\phi_k}{2}+i\frac{\pi_k}{2\omega_k}\hsp
B^\dag_S=\frac{\phi_S}{2}-i\frac{\pi_S}{2\omega_S}\hsp
\frac{B_{-S}}{2\omega_S}=\frac{\phi_S}{2}+i\frac{\pi_S}{2\omega_S}
\eeq
and this basis satisfies
\beq
[B_{k_1},B^\dag_{k_2}]=2\pi \delta(k_1-k_2)\hsp
[B_S,B^\dag_S]=1\hsp [\phi_0,\pi_0]=i. \label{eq:commutation}
\eeq

\sloppy Any operator in the operator algebra may be expressed in either the $\{A^\dag_p,A_p\}$ basis or the $\{B^\dag_k,B_k,B^\dag_S,B_S,\phi_0,\pi_0\}$ basis.  We will define, correspondingly, two normal orderings.  Plane-wave normal-ordering $::_a$ places all $A$ to the right of $A^\dag$.  Normal-mode normal-ordering $::_b$ places all $B$ and $\pi_0$ to the right of all $B^\dag$ and $\phi_0$.  Our defining Hamiltonian and the kink Hamiltonian found above are plane-wave normal-ordered.  However it will be convenient to normal-mode normal-order the kink Hamiltonian.  This can be done using the commutation relations and decompositions above.  

Combining the various decompositions above, one can derive the Bogoliubov transformation that relates the two bases
\bea
A^\dag_p&=&\tg_B(p)\left(\frac{\phi_0}{2}-\frac{i\pi_0}{2\omega_p}\right)+\frac{\tg_S(p)}{2}\left[\left(1+\frac{\os}{\op}\right)B^\dag_S+\left(1-\frac\os\op\right)\frac{B_S}{2\os}\right]\\
&&+
\pin{k}\frac{\tg_k(p)}{2}\left[\left(1+\frac{\ok{}}{\op}\right)B^\dag_k+\left(1-\frac{\ok{}}{\op}\right)\frac{B_{-k}}{2\ok{}}
\right]\nonumber\\
\frac{A_{-p}}{2\op}&=&\tg_B(p)\left(\frac{\phi_0}{2}+\frac{i\pi_0}{2\omega_p}\right)+\frac{\tg_S(p)}{2}\left[\left(1-\frac{\os}{\op}\right)B^\dag_S+\left(1+\frac\os\op\right)\frac{B_S}{2\os}\right]\nonumber\\
&&+
\pin{k}\frac{\tg_k(p)}{2}\left[\left(1-\frac{\ok{}}{\op}\right)B^\dag_k+\left(1+\frac{\ok{}}{\op}\right)\frac{B_{-k}}{2\ok{}}
\right]\nonumber
\eea
where we have defined the inverse Fourier transform
\beq
\tg(p)=\int dx g(x) e^{ipx}.
\eeq
Below we will also encounter the simpler combinations
\bea
A^\dag_p+\frac{A_{-p}}{2\op}&=&\tg_B(p)\phi_0+\tg_S(p)\left[B^\dag_S+\frac{B_S}{2\os}\right]+\pin{k}\tg_k(p)\left[B^\dag_k+\frac{B_{-k}}{2\ok{}}\right]\\
A^\dag_p-\frac{A_{-p}}{2\op}&=&-i\tg_B(p)\frac{\pi_0}{\op}+\frac{\os\tg_S(p)}{\op}\left[B^\dag_S-\frac{B_S}{2\os}\right]+\pin{k}\frac{\ok{}\tg_k(p)}{\op}\left[B^\dag_k-\frac{B_{-k}}{2\ok{}}\right].\nonumber
\eea

%%%%%%%%%%%%%%%%%%%%%%%%%%%%%%%%%%
\subsection{The Kink Hamiltonian} \label{sec:doingexpansion}
%%%%%%%%%%%%%%%%%%%%%%%%%%%%%%%%%%

Now we convert $H_2$ from plane-wave to normal-mode normal-ordering.  The calculation can be found in \ref{appn}.
The result consists of the sum of a $c$-number term $Q_1$, which depends on the normal modes and so the choice of classical solution $\Phi_{x_0}(x)$, and a normal-mode normal-ordered operator $\tilde{H}'_2$
\beq
H\p_2=Q_1+\tilde{H}\p_2.
\eeq
Specifically, 
\beq
Q_1=-\frac{1}{4}\pin{p}\left[
\left|\tg_B(p)\right|^2\op
+\left|\tg_S(p)\right|^2\frac{(\os-\op)^2}{\op}
+\pin{k}\left|\tg_k(p)\right|^2
\frac{(\ok{}-\op)^2}{\op}
\right] \label{q1-a}
\eeq
and the operator is
\beq 
\tilde{H}\p_2=\frac{\pi_0^2}{2}+\os B^\dag_SB_S+\pin{k}\ok{} B^\dag_kB_k. \label{h2p}
\eeq
The first term in (\ref{h2p}) is a nonrelativistic kinetic term for a free particle, in this case the (anti)kink center of mass, where the eigenvalue of $\phi_0$ is identified with the particle's position times the square root of its mass.  The other terms are harmonic oscillators for each normal mode.  Therefore the lowest eigenvalue eigenstate of $H\p_2$ is $\vac_0$ defined by
\beq
\pi_0\vac_0=B_S\vac_0=B_k\vac_0=0. \label{vac0}
\eeq
If there is no zero-mode, for example if $\Phi(x)$ is a vacuum solution, then one obtains the same expressions as above but with all terms containing $g_B$ or $\pi_0$ dropped.  Similarly in the absence of a shape mode, the $g_S$ and $B_S$ terms are dropped.

What is $\vac_0$?  It is an eigenstate of $H\p_2$ and, since $H_1\p=0$ and $H_0\p=Q_0$ is a scalar, it is an eigenstate of $H\p$ expanded out to quadratic order in the fields.   In a perturbative approach to the $H\p$ eigenvalue equation,
\beq
H\p\vac=Q\vac\hsp \text{where} \hspace{.7cm} H\p=\sum_{i=0}^{\infty}H\p_i\hsp\vac=\sum_{i=0}^{\infty}\vac_i\hsp Q=\sum_{i=0}^{\infty}Q_i,
\eeq
$H\p_2$ contributions will be suppressed with respect to those of $H\p_0$ by one power of $\hbar$.  In this sense, $\mathcal{D}_\Phi\vac_0$ is the one-loop approximation to the ($H$ eigenstate) ground state $\mathcal{D}_\Phi\vac$ of the $\Phi(x)$-sector.    This situation is summarized in Table~\ref{eigentab}. In the BPS impurity model \eqref{Lag-imp} the expansion parameter is $\lambda$ and one could rewrite the perturbative expansions as 
\beq
 H\p=\sum_{i=0}^{\infty}\lambda^{i/2} \hat{H}\p_i\hsp\vac=\sum_{i=0}^{\infty}\lambda^{i/2} \hat{\vac}_i\hsp Q=m \sum_{i=0}^{\infty}\lambda^{i-1} \hat{Q}_i
\eeq
where the factors of $\lambda$ have been pulled out of all terms on the right hand sides and $\hat{Q}_i$ are dimensionless.  However we do not adopt this convention.

\begin{table}
\begin{center}
\begin{tabular}{|c|c|c|c|}
Operator&Eigenstate&Interpretation&Eigenvalue\\
\hline
$H_2\p$&$\vac_0$&1-loop ground state of $\Phi(x)$-sector&$Q_1$\\
$H\p$&$\vac$&Ground state of $\Phi(x)$-sector&$Q$\\
$H$&$\mathcal{D}_\Phi \vac$&Ground state of $\Phi(x)$-sector&$Q$\\
\hline
\end{tabular}
\caption{Operators and eigenstates} \label{eigentab}
\end{center}
\end{table}

The topological sector including $\Phi_{x_0}(x)$ may contain other classical solutions.  However, $H\p$ depends on the choice of $\Phi_{x_0}(x)$ via (\ref{df}) and so $\vac_0$ will also depend on this choice.  A different choice would lead to a different $\vac_0$, although presumably, in the sense of an asymptotic series, after performing the perturbative summation one would arrive at the same $H$ eigenstate $\mathcal{D}_\Phi\vac$.

%%%%%%%%%%%%%%%%%%%%%%%%%%%%%%%%%%
\subsection{$\phi_0$ and the Antikink's Position} \label{phi0sez}
%%%%%%%%%%%%%%%%%%%%%%%%%%%%%%%%%%

\begin{table}
\begin{center}
\begin{tabular}{|c|c|c|c|}
Variable&Meaning&Definition\\
\hline
%$x$&Label of a classical kink solution&$\phi(x,t)=\Phi_x(x)$\\
$Q_0$&Classical antikink mass&$H\p_0=Q_0$\\
$Q_1$&One-Loop antikink mass&$H\p_2\vac_0=Q_1\vac_0$\\
$\Phi_{x_0}(x)$&Classical antikink solution at modulus $x_0$&$\phi(x,t)=\Phi_{x_0}(x)$\\
$\mathcal{D}_{\Phi_{x_0}}$&Displacement operator by ${\Phi_{x_0}}$&$\phi(x)\mathcal{D}_{\Phi_{x_0}}=\mathcal{D}_{\Phi_{x_0}}\left(\phi(x)+\Phi_{x_0}(x)\right)$
\\
$x_0$&Point in moduli space where $H\p$ is defined& $H\p=\mathcal{D}_{\Phi_{x_0}}^\dag H\mathcal{D}_{\Phi_{x_0}}$\\
$x_1$&Centroid of antikink wave packet&$\langle y|\mathcal{D}_{\Phi_{x_0}}^\dag \phi(x)\mathcal{D}_{\Phi_{x_0}}|y\rangle\approx\Phi_{x_1}(x)$\\
$y$&Eigenvalue of $\phi_0$&$\phi_0|y\rangle=y|y\rangle$\\
\hline
\end{tabular}
\caption{Wave Packet Notation} \label{wptab}
\end{center}
\end{table}

Let us restrict our attention to the case of BPS antikinks $\Phi(x)=\Phi_{x_0}(x)$.   What does Eq.~(\ref{vac0}) tell us about the ground state?  How is the location of the quantum antikink related to $x_0$?

The algebra of operators is generated by $\phi_0$, $\pi_0$, $B_k$ and $B_k^\dag$.  The first two commute with the others, and generate the canonical algebra.  As a result the space of states factorizes into a tensor product of a representation of the canonical algebra and a representation of the various Heisenberg algebras.  As is customary in quantum mechanics, the representation of the canonical algebra may be expanded in terms of wave functions $\psi:\R\rightarrow\C:y\mapsto \psi(y)$,
\beq
|\psi\rangle=\int dy \psi(y)|y\rangle\hsp \phi_0|\psi\rangle=\int dy y\psi(y)|y\rangle\hsp
 \pi_0|\psi\rangle=-i\int dy \frac{\partial\psi(y)}{\partial y}|y\rangle.
\eeq
Tensoring $|y\rangle$ with the ground state of the oscillators, one arrives at $|y\rangle_0$.  It satisfies
\beq
 \phi_0|y\rangle_0=y|y\rangle_0\hsp
 B_k|y\rangle_0=0
 \eeq
where $k$ ranges over all real numbers and also includes any possible shape mode $S$.  The one-loop ground state $\vac_0$ of $H\p$ is annihilated by $\pi_0$ and so corresponds to a nonnormalizable, flat superposition of all $|y\rangle_0$
\beq
\vac_0=\int dy   |y\rangle_0. \label{flat}
\eeq

%The operators $\phi_0$ and $\pi_0$ satisfy the canonical commutation relations and commute with the rest of the basis of our operator algebra.  Therefore, as in quantum mechanics, we may decompose our states into eigenstates of $\phi_0$ with eigenvalues $y$
%\beq
%\phi_0|y\rangle=y|y\rangle.
%\eeq
%The operator algebra contains $B^\dag$ and $B$ and so the label $y$ does not completely specify the state, there will also be other quantum numbers which will not play a role in present discussion\footnote{Some of the arguments here will be repeated in Sec.~\ref{smearsez}, where the other quantum numbers are included.}.  Just as in quantum mechanics, states may be expressed in terms of wave functions, or equivalently as linear combinations of our basis states
%\beq
%|\psi\rangle=\int dy \psi(y)|y\rangle.
%\eeq

How is $y$ related to the moduli space of classical solutions of Eq.~(\ref{feq})?  Let $x_0$ be a point on the classical moduli space corresponding to the classical position of the solution $\Phi_{x_0}(x)$.  
Then, in the state $\mathcal{D}_{\Phi_{x_0}}|y\rangle_0$, the expectation value of $\phi(x)$ is
\bea
\frac{{}_0\langle y|\mathcal{D}_{\Phi_{x_0}}^\dag \phi(x)\mathcal{D}_{\Phi_{x_0}}|y\rangle_0}{{}_0\langle y|\mathcal{D}_{\Phi_{x_0}}^\dag \mathcal{D}_{\Phi_{x_0}}|y\rangle_0}&=&\frac{{}_0\langle y|(\phi(x)+\Phi_{x_0}(x))|y\rangle_0}{{}_0\langle y|y\rangle_0}=\Phi_{x_0}(x)+g_B(x){}\frac{_0\langle y|\phi_0|y\rangle_0}{{}_0\langle y|y\rangle_0}=\Phi_{x_0}(x)+y g_B(x)\nonumber\\
&=&\Phi_{x_0}(x)+ \frac{y}{\sqrt{M(x_0)}} \partial_{x_0} \Phi_{x_0}(x)=\Phi_{x_0+y/\sqrt{M(x_0)}}(x)+O(y^2).
 \label{espa}
\eea
So while one might be tempted to identify $\Phi_{x_0}(x)$ with the expectation value of the scalar field $\phi(x)$, this is not quite right.  At small $y$, the expectation value approaches $\Phi_{x_1(y)}(x)$ where
\beq
x_1(y)=x_0+\frac{y}{\sqrt{M(x_0)}}. \label{x1def}
\eeq

%Let us try to do a bit better.  Recall that $g_B(x)$ is proportional to $\partial_{x_0}\Phi_{x_0}(x)$.  This means that for each eigenvalue $y$, one may associate the point $x_1(y)$ in the classical moduli space according to the definition
%\beq
%\langle y|\mathcal{D}_{\Phi_{x_0}}^\dag \phi(x)\mathcal{D}_{\Phi_{x_0}}|y\rangle=\Phi_{x_1}(x)+O(y^2). \label{x1def}
%\eeq
%Here the $y g_B(x)$ term in (\ref{espa}) is the linear term in a power series expansion of $\Phi_{x_1}(x)$ with respect to $x_1$ about the point $x_1=x_0$
%\beq
%\Phi_{x_1}(x)=\Phi_{x_0}(x)+(x_1-x_0)\partial_{x_0}\Phi_{x_0}(x)+O\left((x_1-x_0)^2\right)\hsp
%y=\frac{x_1-x_0}{\sqrt{M(x_0)}} \label{power}
%.
%\eeq
At $y=0$ one immediately sees that $x_1(0)=x_0$.  This is the zeroeth order term in the expansion (\ref{espa}).  On the other hand, this expansion is meaningless at large $y$.  However, for small $y$, it tells us that {\it{$x_1(y)$ is the expected position of the quantum antikink in the classical moduli space.}}  

It is unsurprising that $x_0$ is not the expected position, as $x_0$ was chosen arbitrarily.  However, $x_0$ must be chosen close to $x_1$ if $y$ is to be small, and so (\ref{x1def}) usefully constrains $x_1$.   It may appear that the definition (\ref{x1def}) of $x_1$ also depends on $x_0$.  Our point of view will be as follows.  One first fixes a localized quantum state corresponding to a kink smeared about some point $x_1$ in the classical moduli space.  Then, for any given choice of $x_0$, one will see that the eigenvalues $y$ of $\phi_0$ that dominate $\psi(y)$ depend on both $x_0$ and $x_1$.  In this sense, the observable $x_1$ is independent of the arbitrary choice of $x_0$.  

Summarizing, we have two distinct sets of coordinates which describe the position of the quantum antikink in the classical moduli space.  First, the coordinates $x_1$ yield the classical solution $\Phi_{x_1}(x)$ closest to the expectation value of the scalar field.  Second, if we fix $x_0$, then the eigenvalue $y$ of $\phi_0$ provides a second set of coordinates for the antikink position.  With $x_0$ fixed, these two coordinates on the moduli space are related by the function $x_1(y)$.  %Let us investigate this function further.

\begin{figure}[h] %[!tph]
	\begin{center}
		%\includegraphics[width=3in]{V2s.pdf}
		%\caption{The potential $V^{2}[f_{\phi_0}]$ for a variety of moduli $\phi_0 \in [0,5]$.}
\includegraphics[width=4in]{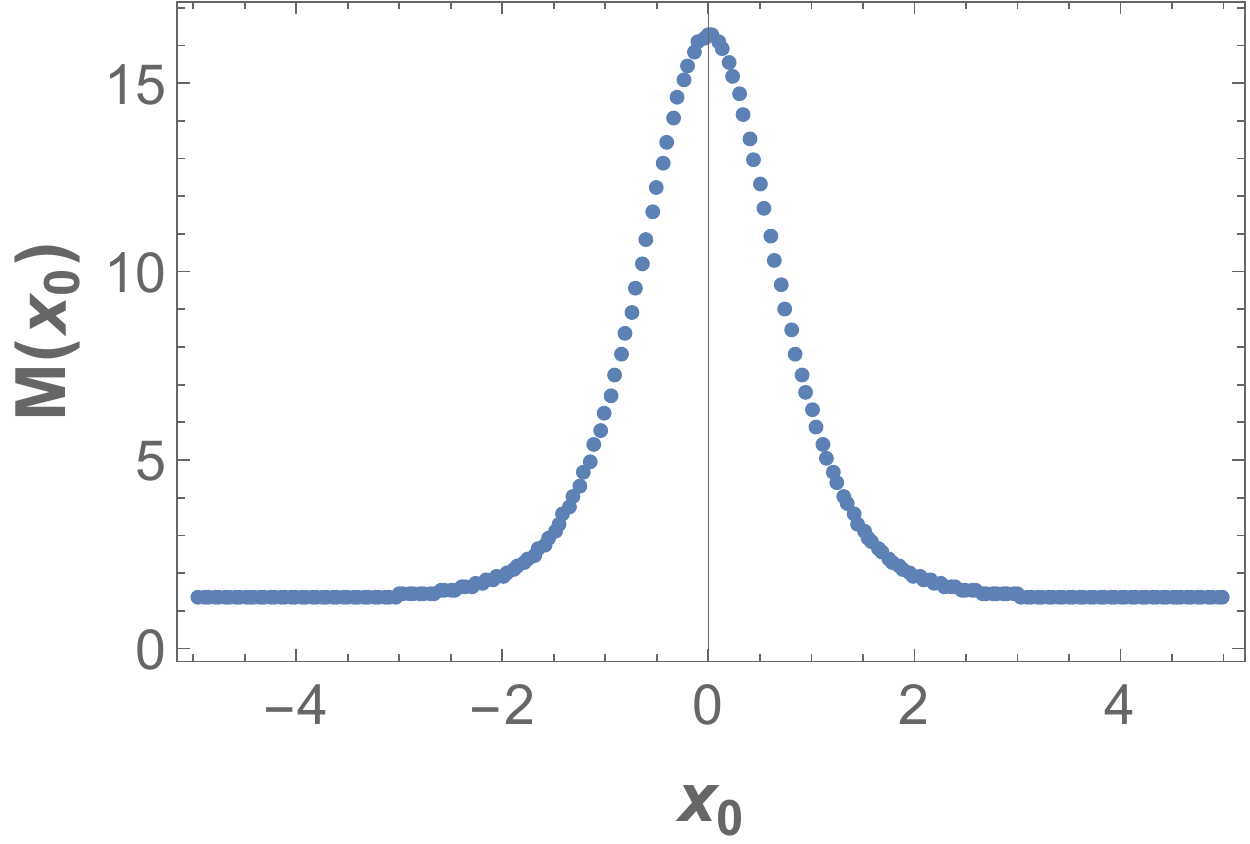}
		\caption{The moduli space metric $M(x_0)$ evaluated numerically at $\alpha=3$, $m=2$, $\lambda=l=1$.  Notice that, at $|x_0|>3$ the antikink and the impurity are well separated and the expression tends to $Q_0=4/3$, in agreement with Eq.~(\ref{q0m}).}
\label{repfig}
	\end{center}
\end{figure}

%Recall that the zero-mode $g_B(x)$ is proportional to the derivative $\partial \Phi_{x_0}/\partial x_0$ and it is normalized as in (\ref{gnorm}).  Therefore, combining Eqs.~(\ref{espa}) and (\ref{x1def}), one finds
%\beq
%\left|\frac{\partial x_1(y)}{\partial y}\right|=\sqrt{M(x_0)} +O(y)\label{x1def}
%\eeq
%which is independent of $x$.  For example, 
In the case of a BPS antikink with no impurity, $\Phi_{x_0}(x)$ is a function only of $x-x_0$ and so
\beq
\sqrt{M(x_0)}g_B(x)=\frac{\partial \Phi_{x_0}(x)}{\partial{x_0}}=-\frac{\partial \Phi_{x_0}(x)}{\partial {x}}.
\eeq
Inserting this into the BPS condition
\beq
Q_0=\int dx \left(\frac{\partial \Phi_{x_0}(x)}{\partial {x}}\right)^2=M(x_0)\int dx g_B^2(x)=M(x_0). \label{q0m}
\eeq
Therefore we find that in the absence of an impurity, the moduli space metric $M(x_0)=M_0$ is simply the classical kink mass $Q_0$. The impurity modified $M(x_0)$ is plotted in Fig.~\ref{repfig}.  As expected from (\ref{q0m}), when $|x_0|$ is sufficiently large, so that the antikink and impurity are well separated, the function tends to $Q_0=2m/3\lambda=4/3$.

%which, inserted into (\ref{x1def}) using (\ref{gnorm}), yields
%\beq
%\left|\frac{\partial x_1(y)}{\partial y}\right|=\frac{1}{\sqrt{Q_0}}. \label{purerep}
%\eeq
%In the case with no impurity we conclude that there is a simple linear relationship $x_1=x_0\pm y/\sqrt{Q_0}$ between the two sets of coordinates, $x_1$ and $y$, on the classical moduli space.  We will see that this is also the case when the antikink is sufficiently far from an impurity that the normal mode spectrum changes slowly with respect to the modulus.

%Recall that $H_2\p$ contains the nonrelativistic kinetic energy term $\pi_0^2/2$ which led us to identify the eigenvalue of $\phi_0$ is the square root of the inertial kink mass times the position. 

More generally, Eq.~(\ref{x1def}) relates the classical moduli space, parametrized by $x_1$, to the eigenvalue $y$ of the operator $\phi_0$.  In the example treated below $x_1$ will be a global coordinate on the moduli space, however since $y$  is only defined once $x_0$ is fixed, $y$ will be in a neighborhood of $0$ with a value depending on $x_0$.  One may apply this construction to a general multidimensional moduli space, in which case $y$ is a point in the total space of the tangent bundle of the moduli space, in the fiber over the point $x_0$.  

When the eigenvalues $y$ of $\phi_0$ are large, the corrections at higher loops will also be large.  These corrections are of course suppressed by powers of the coupling, and so the semi-classical expansion can only converge in the sense of an asymptotic series for eigenvalues which are much smaller than order $1/\sqrt{Q_0}$.

This leads us to a complication.  The equation $\pi_0\vac_0=0$ means that the wave function $\psi(y)$ is independent of $y$, and so eigenvectors $|y\rangle_0$ with all eigenvalues $y$ appear in the ground state (\ref{flat}).  The states are in fact dominated by eigenvalues beyond the range in which our perturbative expansion of the states is valid.  For a translation-invariant Hamiltonian, such as one with no impurity, this is a problem for finding the states but not for finding the energy $Q$ since translation-invariance implies that the energy is anyway independent of the antikink location.  However in the case of an impurity, we will see that $Q_1$ depends on the value of $x_0$ even within a classically-degenerate moduli space.  This will prevent us from finding states that are Hamiltonian eigenvectors, even in the sense of an asymptotic series.  Indeed, in the example treated below we will see that the impurity is repulsive and so, just like the case of a particle in a repulsive potential in quantum mechanics, there are no reasonable eigenstates.

Our goal is to study the dynamics of quantum antikinks near an impurity.  For this we are interested in antikinks that are spatially localized, not in exact Hamiltonian eigenstates, which would anyway be stationary and so have no dynamics.  This motivates us to study spatially localized wave packets, centered at some small eigenvalue $y=y_1$.  We will always take the width to be sufficiently small that the support of the wave packet is within the perturbative regime, avoiding the problem described above.  

%%%%%%%%%%%%%%%%%%%%%%%%%%%%%%%%%%
\subsection{One-Loop Energy} \label{sec:oneloop}
%%%%%%%%%%%%%%%%%%%%%%%%%%%%%%%%%%

Summarizing, we will not study exact Hamiltonian eigenstates, but rather spatially localized wave packets.  These wave packets are centered on some value $y_1$ of $y$.  We have argued that each value $y_1$ corresponds to a classical solution $\Phi_{x_1(y_1)}(x)$, and so equivalently each wave packet describes configurations close to some classical solution.

Now our perturbative expansion will involve a power series in $y_1$ which, for a given wave packet, is most convergent if we fix $x_0=x_1$ so that $y_1=0$.   With this choice, $Q_1$ depends on $\Phi_{x_0}(x)$ and so on $x_0$.  If there is no impurity, then translation-invariance guarantees that any choice of modulus leads to the same $Q_1$ and so this initial choice of $x_0$ does not affect the one-loop energy.

With this caveat in mind, the energy of our wave packet state is simply given by $Q_1(x_0)$, with $x_0$ chosen near the center $x_1$ of the wave function, and the width of the energy depends on the variation in $Q_1(x_0)$ over the width of the wave function $\psi(y)$.  The energy $Q_1$ is given by summing the $c$-number contributions from Subsec.~\ref{contsez}
\beq
Q_1(x_0)=-\frac{1}{4}\pin{p}\left[
\left|\tg_B(p)\right|^2\op
+\left|\tg_S(p)\right|^2\frac{(\os-\op)^2}{\op}
+\pin{k}\left|\tg_k(p)\right|^2
\frac{(\ok{}-\op)^2}{\op}
\right]. \label{q1}
\eeq
We recognize this as the Cahill, Comtet, Glauber formula \cite{cahill76,memassa} for the mass of a (anti)kink in the absence of an impurity. We note that the normal modes and their frequencies also implicitly depend upon $\Phi_{x_0}(x)$ and hence upon $x_0$.

The interpretation now is somewhat different from that of Ref.~\cite{cahill76}.  The mass of a (anti)kink is the difference between the ground state energy of the (anti)kink sector and the vacuum sector.  $Q_1$ is the one-loop energy of the ground state of the $\Phi(x)$-sector.  In the absence of an impurity, as a result of the plane-wave normal-ordering of the defining Hamiltonian, the vacuum energy would vanish at this order.  However, in the presence of an impurity, it is not even guaranteed that there are any Hamiltonian eigenstates in the vacuum sector.  Of course, given a choice of $\sigma(x)$ one can determine whether the vacuum sector has a ground state and calculate its energy using old fashioned perturbation theory.  This energy is independent of the modulus $x_0$ that we choose for our antikink, and so the result of this calculation will simply subtract a $x_0$-independent constant from $Q_1(x_0)$ calculated using the BPS antikink solution $\Phi_{x_0}(x)$.  Therefore the antikink mass is, up to corrections resulting from the width of the wave packet, equal to $Q_1(x_0)$ plus a $x_0$-independent constant.  

This means that the energy equivalence of BPS solitonic solutions is lifted at the one-loop quantum order. Classically, the BPS (anti)kink in our BPS-impurity model can be located at any distance from the impurity. In other words, there is no static force between the soliton and impurity. This is no longer true if the quantum corrections are taken into account. Depending on its position, $x_1$, the energy of the soliton-impurity system changes leading to appearance of a force.  This is no surprise, such a lifting of the classical degeneracy has already been observed in a two field model~\cite{deglift}.

The derivation above may be repeated for the one-loop correction to the energy of any state.  In some cases, such as a vacuum sector in a model with possible impurities, the moduli space contains only discrete classical solutions.  In that case the derivation proceeds as above except there is no zero mode in the decomposition of the fields.   Completing the derivation as above, one finds that in such a case the energy is given by (\ref{q1}) without the zero mode term $\tg_B$. In \ref{vacsez} we present the calculation in the vacuum sector, with an approximate analytical result.

%%%%%%%%%%%%%%%%%%%%%%%%%%%%%%%%%%
\subsection{The BPS impurity $\phi^4$ model} \label{kinksez}
%%%%%%%%%%%%%%%%%%%%%%%%%%%%%%%%%%
Now we analyze the system introduced in Sec. \ref{sec:imp}, which is a BPS-impurity deformation of the well-known $\phi^4$ model. The impurity is chosen in the form (\ref{imp-exm}). As described earlier, fluctuations around the antikink $\{g_i\}$ satisfy the Sturm-Liouville equation
\begin{equation} \label{SLfluc}
-\frac{d^2g_i(x)}{dx^2} + \frac{m^2}{2}(3\lambda\Phi_{x_0}(x)^2 - 1 - \sqrt{2}\alpha  \sech(lx)^2)g_i(x) = \omega_i^2 g_i(x).
\end{equation}
We solve \eqref{SLfluc} numerically for $m=2$, $\lambda=l=1$ with a weak ($\alpha=0.3$) and strong ($\alpha=3$) impurity. This is performed in two ways, which are described in \ref{Num}.  The resulting mode structure is shown in Fig. \ref{omfig}. Then, we calculate $Q_1(x_0)$ using \eqref{q1}.

\begin{figure} %[!tph]
	\begin{center}
		\includegraphics[width=0.35\textwidth]{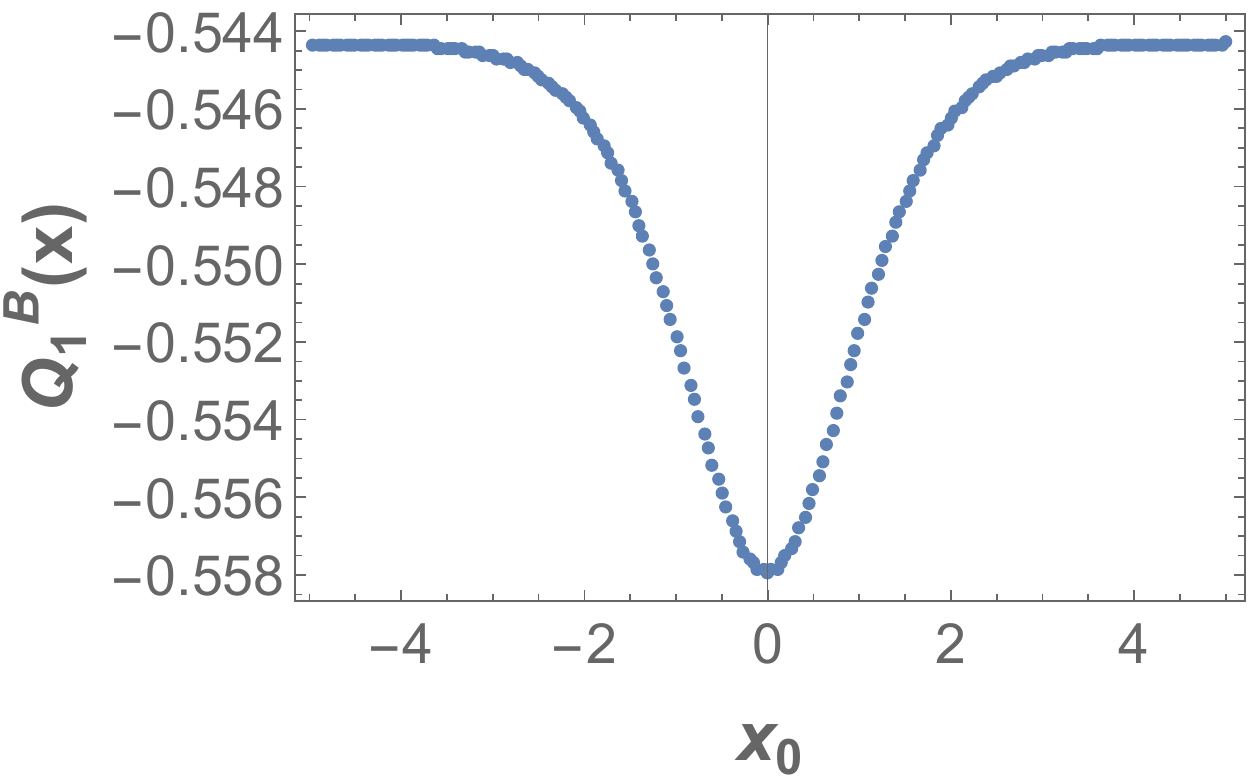}
		\includegraphics[width=0.35\textwidth]{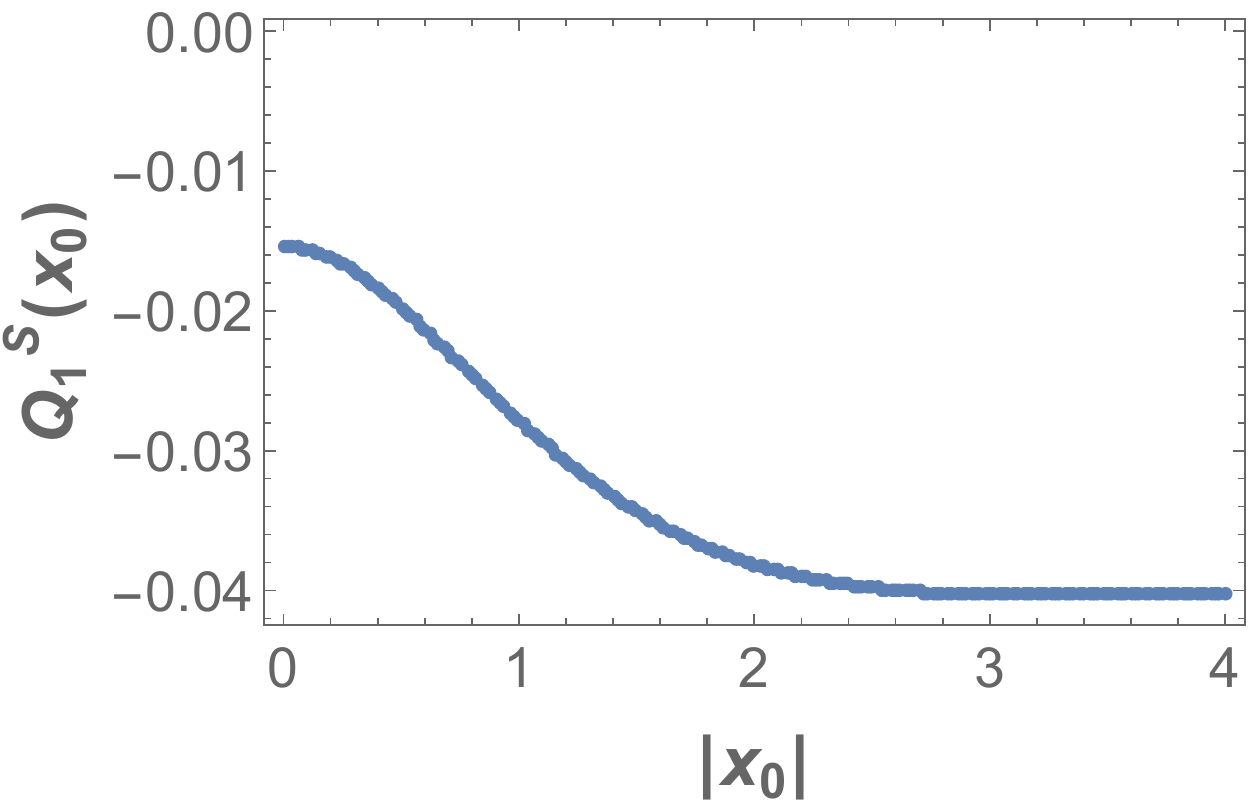}
\includegraphics[width=0.35\textwidth]{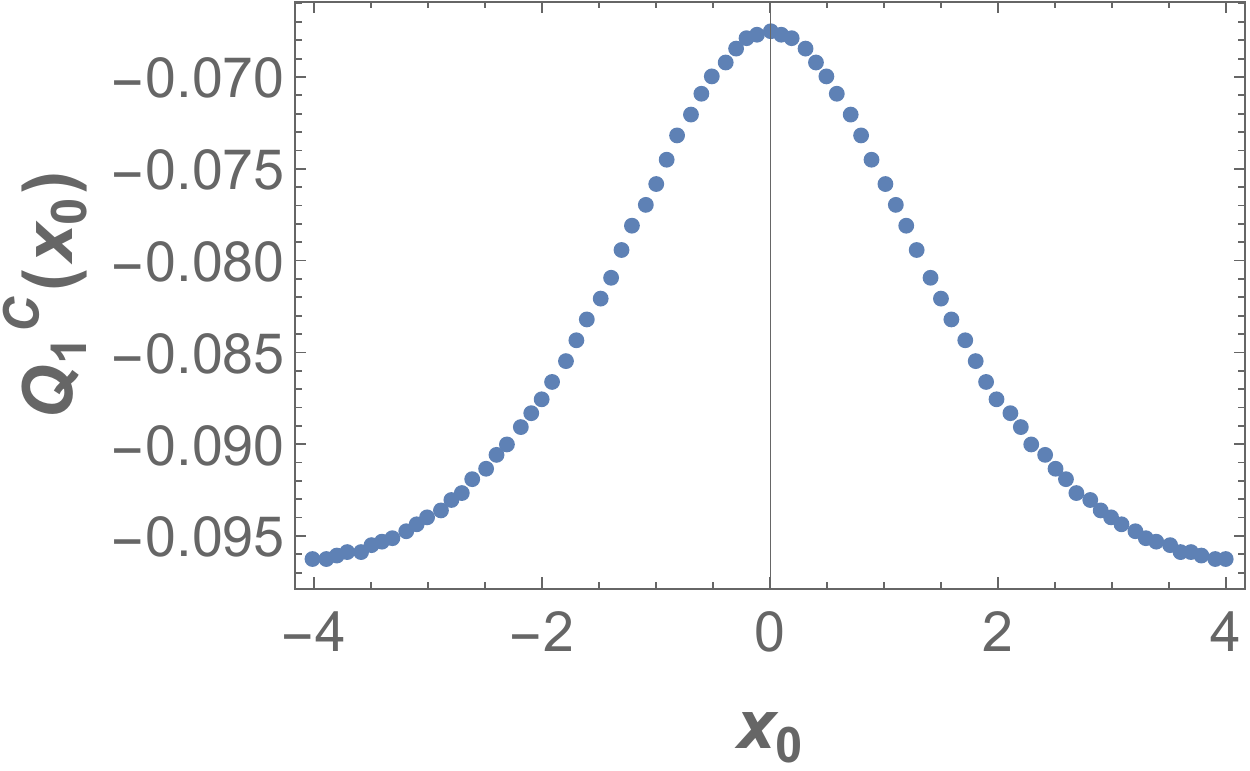}
\includegraphics[width=0.35\textwidth]{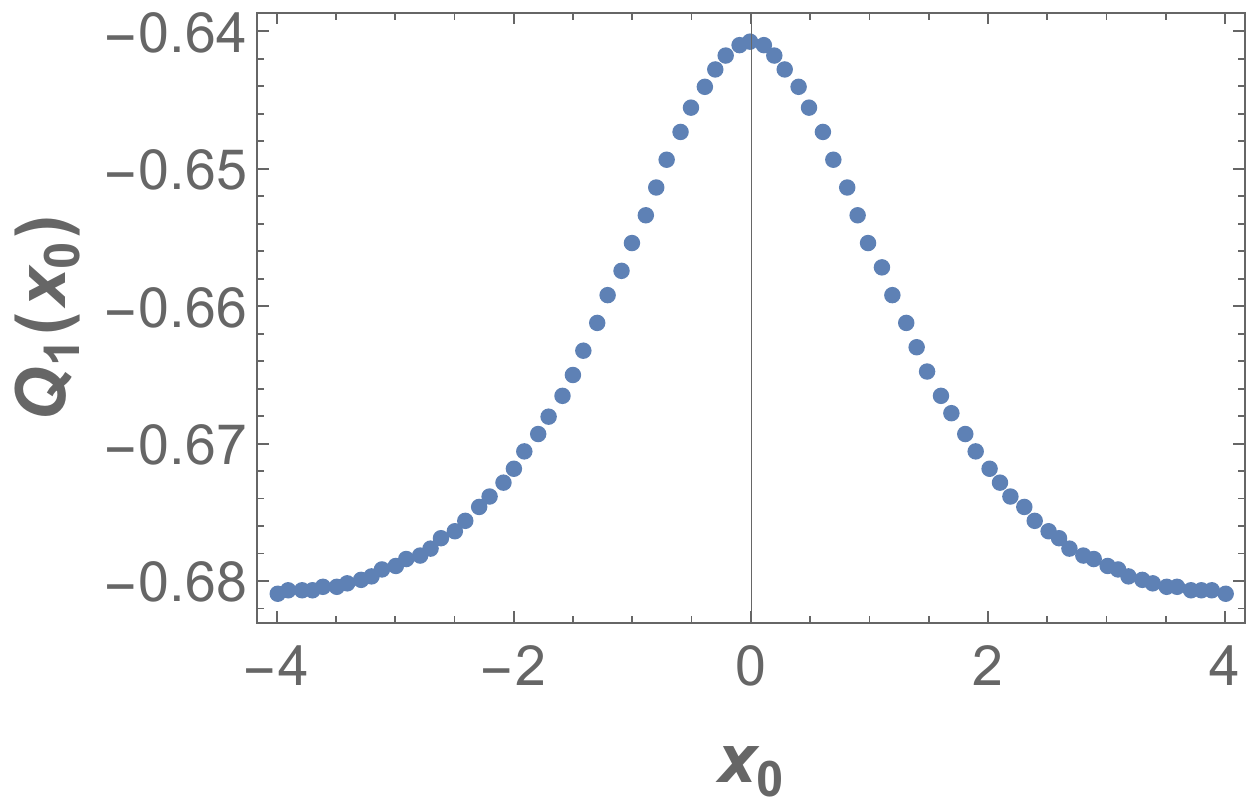}
		\caption{The zero mode (top-left), shape mode (top-right) and continuum (bottom-left) contributions to the one-loop mass correction $Q_1$ (bottom-right) as a function of $x_0$ for $\alpha=0.3$.  Note that it asymptotes to the sum of the vacuum energy ($-0.016$) in this model plus the one-loop mass ($-0.666$) in a model with no impurity \cite{dhn2}.}
		\label{fig:Q1_03}
	\end{center}
\end{figure}
\begin{figure} %[!tph]
	\begin{center}
		\includegraphics[width=0.35\textwidth]{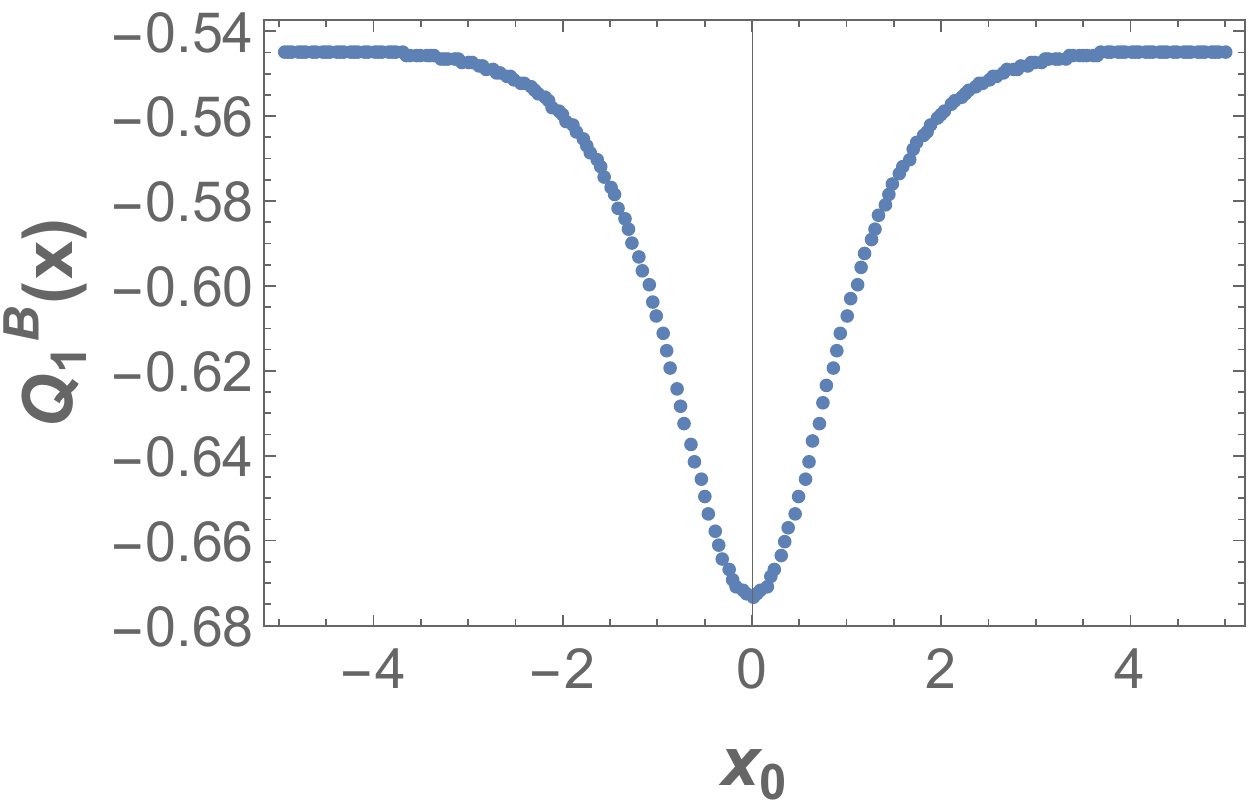}
		\includegraphics[width=0.35\textwidth]{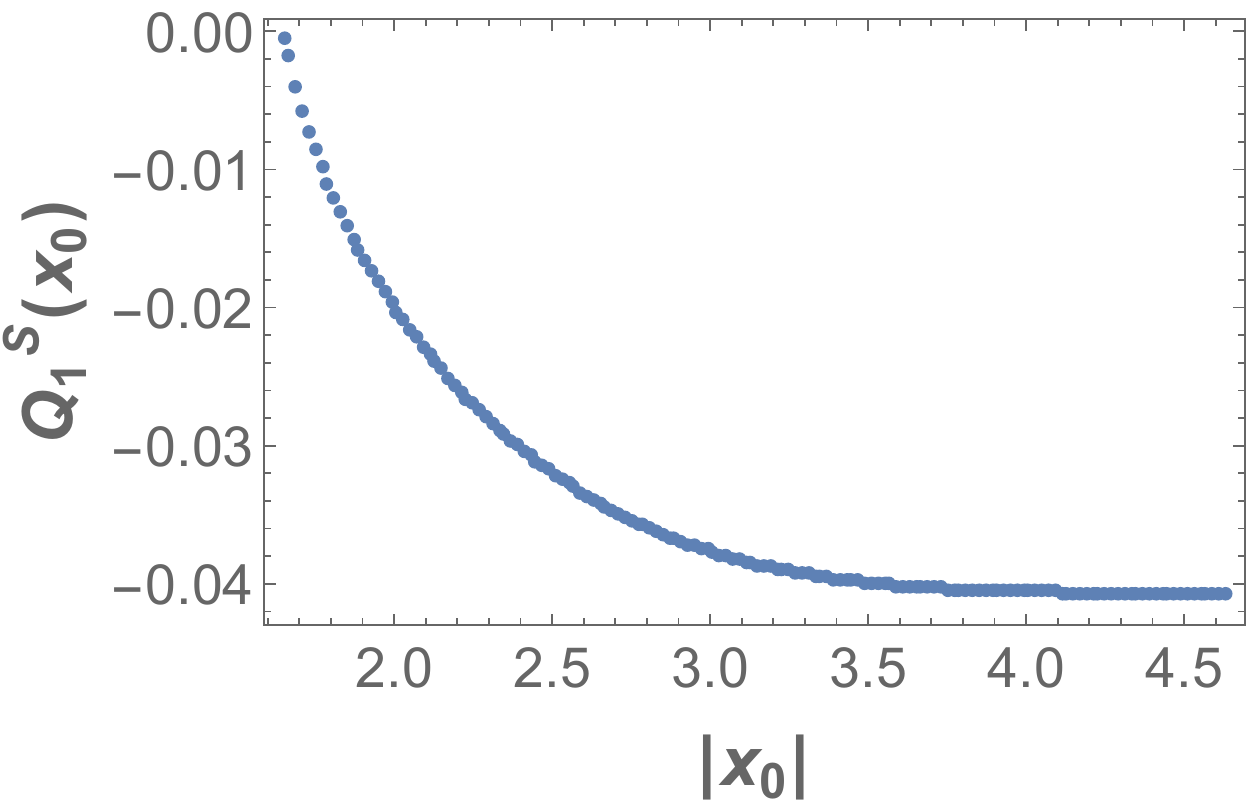}
\includegraphics[width=0.35\textwidth]{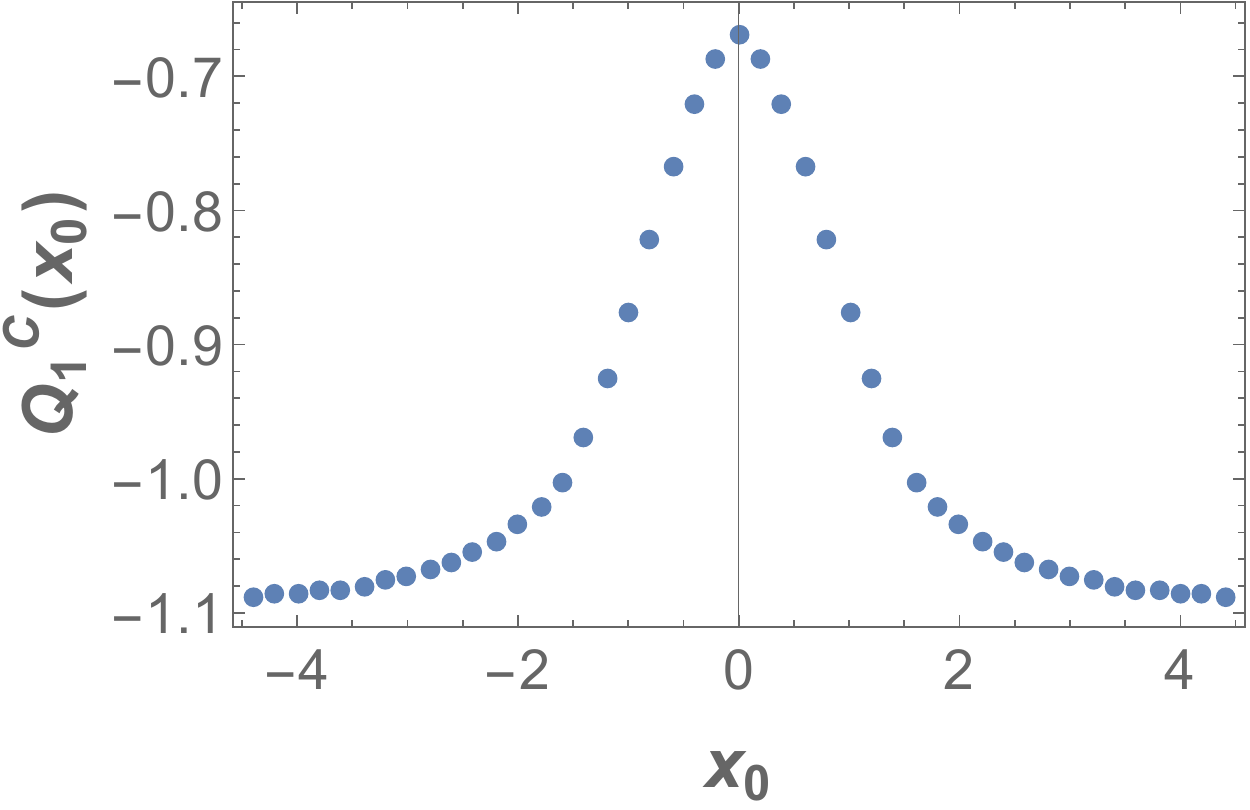}
\includegraphics[width=0.35\textwidth]{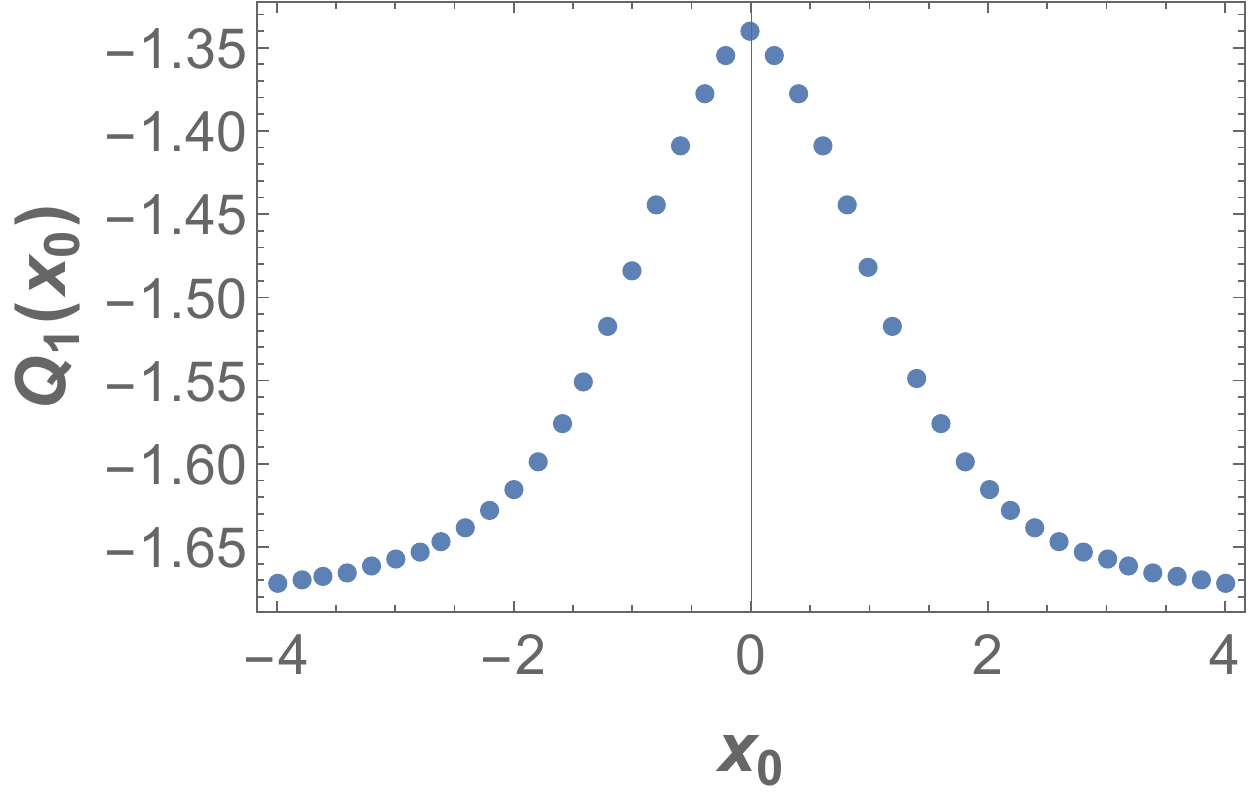}
		\caption{The zero mode (top-left), shape mode (top-right) and continuum (bottom-left) contributions to the one-loop mass correction $Q_1$ (bottom-right) as a function of $x_0$ for $\alpha=3$.  Note that it asymptotes to the sum of the vacuum energy ($-1.0$) plus the one-loop mass ($-0.666$) in a model with no impurity \cite{dhn2}.}
		\label{fig:Q1_3}
	\end{center}
\end{figure}
\noindent

For a moment, let the $\alpha$-dependence of $Q_1$ be explicit.   Then $Q_1(x_0, \alpha)$ is the one-loop correction as a function of the modulus $x_0$ and the impurity strength $\alpha$. There are some consistency checks: $Q_1(0,0) = -m/3=-2/3$, the one-loop antikink correction to the long-understood~\cite{dhn2} $\phi^4$ model and $Q_1(\infty,\alpha) = Q_1(0,0) + Q_1^\text{vac}$, saying that when the antikink is far from the impurity their contributions decouple.

We plot $Q_1(x_0, 0.3)$  in Fig.~\ref{fig:Q1_03} and $Q_1(x_0, 3.0)$  in Fig.~\ref{fig:Q1_3}. Note that the consistency checks are satisfied, showing that the numerical methods presented in \ref{Num} appear to work. The two methods also give the same result, confirming consistency between them. Overall, we see that the classical degeneracy is lifted and the BPS property is destroyed by quantum corrections. The energy is lower when the antikink and impurity are separated, suggesting that the interaction is repulsive. The size of the correction and the difference between vacuum and antikink masses both grow with $\alpha$, as expected.

\begin{figure} %[!tph]
	\begin{center}
		\includegraphics[width=0.35\textwidth]{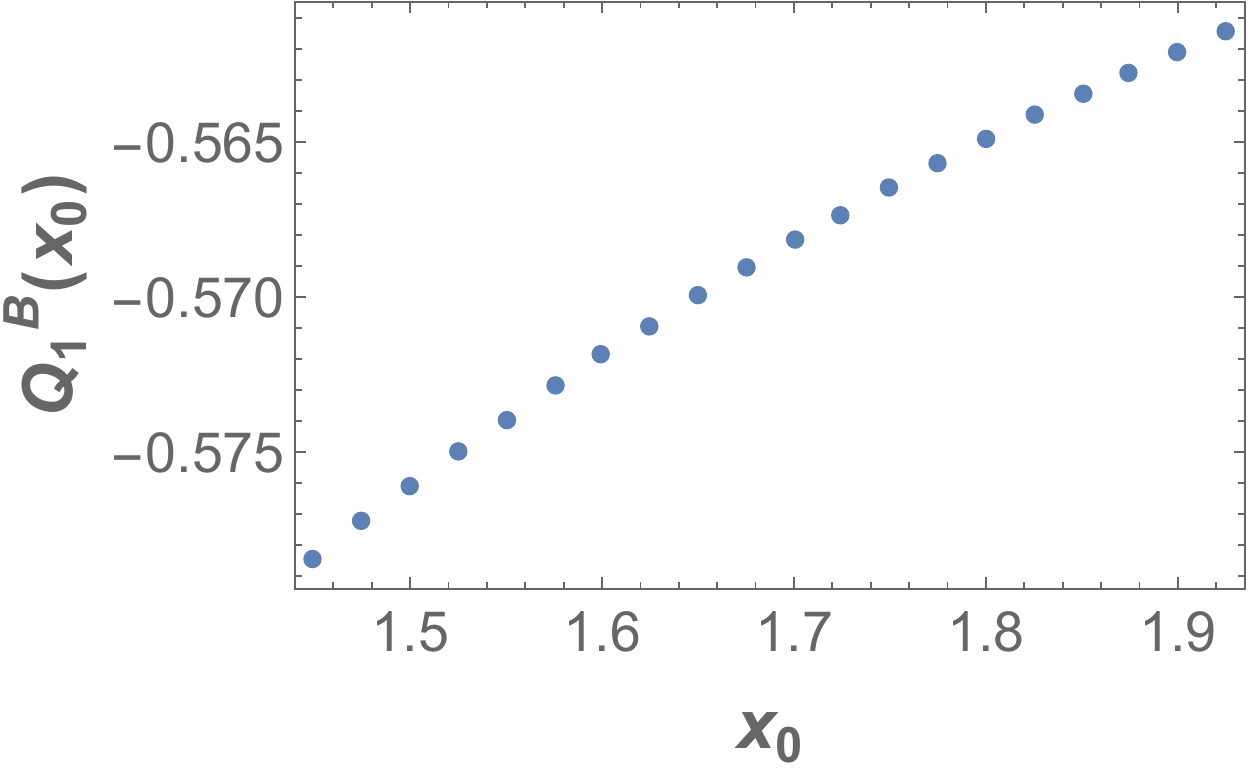}
		\includegraphics[width=0.35\textwidth]{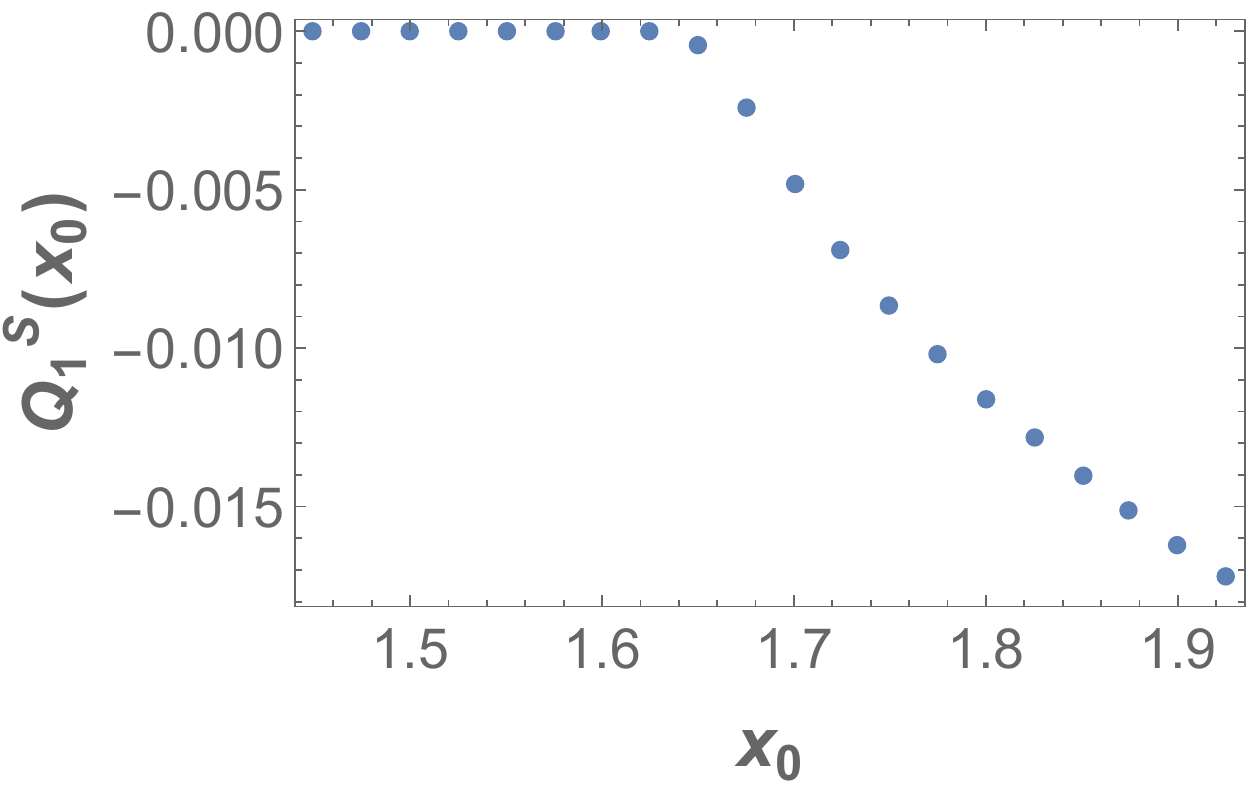}
\includegraphics[width=0.35\textwidth]{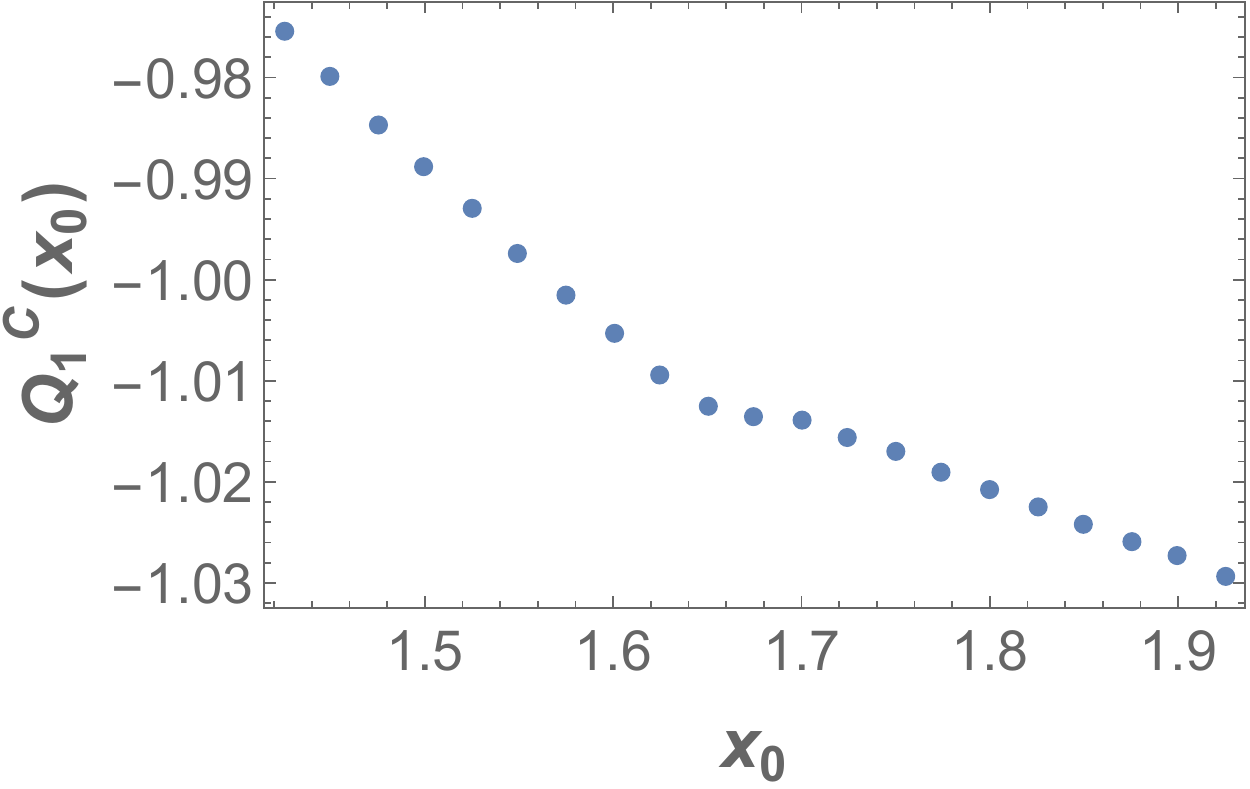}
\includegraphics[width=0.35\textwidth]{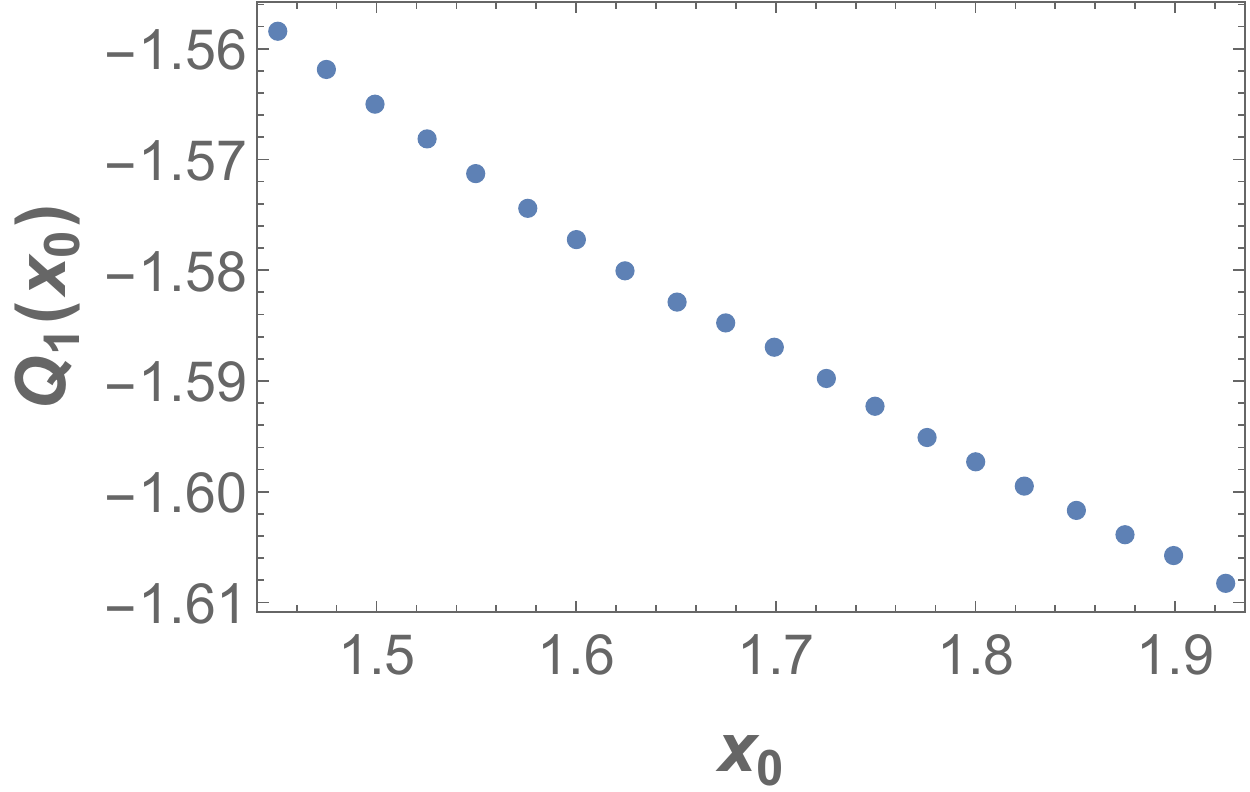}
		\caption{As in Fig.~\ref{fig:Q1_3} but zoomed in close to the wall.  One sees that the shape mode contribution $Q_1^S$ has a nonvanishing derivative at the wall, and so yields a nondifferentiable contribution to $Q_1$.  However the continuum contribution $Q_1^C$ has a corner that, to within the precision of our computation, cancels the corner in $Q_1^S$.  As a result we find no evidence that $Q_1$ fails to be differentiable, and so the force exerted by the impurity appears to be continuous at the wall.}
		\label{fig:Q1zoom}
	\end{center}
\end{figure}
\noindent

In Sec. \ref{accsez} we will show that the force exerted on the antikink by the impurity is proportional to $Q_1\p$ and so the derivative of $Q_1$ is physically important. Since the classical dynamics are dramatically altered near the spectral wall, it is worth looking at $Q_1$ and $Q_1'$ carefully near this point.  In Fig.~\ref{fig:Q1zoom} we show the various contributions to $Q_1$ close to the $\alpha=3$ wall.  We find a corner in the shape mode contribution $Q_1^S$, leading to a discontinuous contribution to the force $-Q_1\p$.  However, there is a corresponding corner\footnote{Such a corner may be expected as, outside the wall, the continuum modes are restricted by the condition that they must be orthogonal to the shape mode while inside the wall there is no such condition.} in the continuum contribution $Q_1^C$ which cancels that of the shape mode within our numerical accuracy. Thus our results are consistent with a differentiable energy $Q_1$, and so a continuous force $-Q_1\p$.

%%%%%%%%%%%%%%%%%%%%%%%%%%%%%%%%%%
\section{Wave Packets and Energy Smearing} \label{smearsez}
%%%%%%%%%%%%%%%%%%%%%%%%%%%%%%%%%%

In Sec.~\ref{unloopsez} we argued that, in the case of a Hamiltonian without translation invariance, the Cahill-Comtet-Glauber formula (\ref{q1}) for the one-loop correction to the antikink energy depends on an {\it{a priori}} arbitrary choice of modulus $x_0$ and, as a result, is not part of a convergent expansion of the energy of a Hamiltonian eigenstate.  Rather, the interpretation is that it should be thought of as a potential $Q_1(x_0)$ in the moduli space.  We explained that we were interested in antikink wave packets which are localized in moduli space, in the sense that the $x$-intercepts $x_0$ of the corresponding classical solutions $\Phi_{x_0}(x)$ are supported in some small region.  

 In Sec.~\ref{kinksez} we numerically calculated $Q_1(x_0)$.  It is the goal of the present section to estimate the size of the energy smearing and to determine when $Q_1(x_0)$ is a reasonable proxy for the energy of a wave packet localized near $x_0$.  These wave packet states are not Hamiltonian eigenstates, nor even eigenstates of $H\p_2$, because they are not annihilated by $\pi_0$.  Instead they have an energy width arising from two sources. First, from the spread of values of $Q_1(x_0)$ in the range of $x_0$ in the wave packet, leading to a smeared position. Second, from the $\pi_0^2$ term in Eq.~(\ref{h2p}) for $H_2\p$; this does not vanish for a wave packet state, leading to a smeared momentum. 

%We begin by repeating the general arguments of Subsec.~\ref{phi0sez} more carefully, including an explicit choice of wave packet. 

For any positive $\wvname$ define the wave packet 
\beq
|\wvname\rangle=\frac{1}{\pi^{1/4}\sqrt{\wvname}}\int dy \exp{-\frac{(y-y_1)^2}{2\wvname^2}}|y\rangle_0. \label{s0}
\eeq
Fixing the normalization condition
\beq
{}_0\langle y_1|y_2\rangle_0 = \delta(y_1-y_2)
\eeq
one finds that the wave packet satisfies
\beq
\langle\wvname|\wvname\rangle=1\hsp
\langle\wvname|\phi_0|\wvname\rangle=y_1. \label{cent}
\eeq
Summarizing, we have two kinds of coordinates on our one-dimensional classical moduli space, $x$ and $y$,  with an arbitrary base point $x_0$ ($y=0$ in the $y$ coordinates) and an antikink wave packet centered at $x_1(y_1)$ ($y=y_1$ in the $y$ coordinates).  This situation is summarized in Table~\ref{xytab}.  Following (\ref{espa}) and using (\ref{cent}) one can show that the wave packet (\ref{s0}) has the property
\beq
\langle\wvname|\df^\dag \phi(x)\df|\wvname\rangle=\Phi_{x_1(y_1)}(x)+O(y_1^2). \label{esp}
\eeq
Therefore the wave packet is centered at $x_1(y_1)$.
%In the absence of an impurity or for an antikink well-separated from the impurity, $g_B(x)=\pm\Phi_{x_0}\p(x)/\sqrt{Q_0}$ and so (\ref{esp}) consists of the first two terms in a power series expansion of $\Phi_{x_0}(x\pm y_1/\sqrt{Q_0})$ with respect to $x$.  Using the symmetry (\ref{tsim}), which holds far from an impurity, this is equal to $\Phi_{x_1}(x)$ if we define the kink position $x_1=x_0\mp y_1/\sqrt{Q_0}$.   More generally, in the presence of an impurity $g_B(x)$ is proportional to the derivative of the antikink solution $\Phi_{x_0}(x)$ with respect to the classical modulus $x_0$ and so (\ref{esp}) is still a power series of $\Phi_{x_1}(x)$ but with $x_1$ given by (\ref{x1def}).  In either case, the expression (\ref{esp}) is equal to $\Phi_{x_1}(x)$ to linear order in $y_1$ (or equivalently in $x_1-x_0$), leading to the interpretation of $x_1$ as the position of the quantum antikink.
%, and so $y_1/\sqrt{Q_0}$ parametrizes this position modulus, and is equal to the position of the center of the kink when the kink is far from the impurity.  In this sense, $\sigma/\sqrt{Q_0}$ is the width of the wave packet.  

\begin{table}
\begin{center}
\begin{tabular}{|c|c|c|c|}
Variable&Meaning&Definition\\
\hline
%$x$&Label of a classical kink solution&$\phi(x,t)=\Phi_x(x)$\\
$x_0$&Point in moduli space where $H\p$ is defined&$\phi(x,t)=\Phi_{x_0}(x)$, $H\p=\mathcal{D}_{\Phi_{x_0}}^\dag H\mathcal{D}_{\Phi_{x_0}}$\\
$x_1(y_1)$&Centroid of antikink wave packet&$\langle y|\mathcal{D}_{\Phi_{x_0}}^\dag \phi(x)\mathcal{D}_{\Phi_{x_0}}|y\rangle\approx\Phi_{x_1}(x)$\\
$y$&Eigenvalue of $\phi_0$&$\phi_0|y\rangle_0=y|y\rangle_0$\\
$y_1$&Centroid of wave packet in $y$ coordinates &$|\wvname\rangle\sim\int dy \exp{-\frac{(y-y_1)^2}{2\wvname^2}}|y\rangle_0$\\
\hline
\end{tabular}
\caption{Various coordinates on moduli space} \label{xytab}
\end{center}
\end{table}

We can now consider the contribution to the energy width arising from the smeared position  and momentum. First, consider the smeared position. The state $\wvname$ is not an eigenvector of $\phi_0$, but rather it has eigenvalues $y$ smeared with a standard deviation of
\beq
\Delta y=\wvname.
\eeq
Each eigenvalue of $y$ corresponds to a classical modulus $x_1(y)$ according to the definition (\ref{x1def}).  Therefore the standard deviation of the modulus $x_1$ is
\beq
\Delta x_1=\left.\Delta y \frac{\partial x_1(y)}{\partial y}\right|_{y=y_1}=\frac{\wvname}{\sqrt{M(x_0)}}.%\left.\frac{\partial x_1(y)}{\partial y}\right|_{y=y_1}.  
\eeq
Our one-loop estimate, based on the base point $x_0$, of the antikink mass is $Q_1(x_0)$.  We recall that this estimate is most reliable when $y=0$ or equivalently $x_0=x_1$, where we expect the fastest convergence of the semiclassical expansion.  Therefore the derivative of the one-loop antikink mass with respect to the modulus is $\partial Q_1(x_0)/\partial x_0$.  This smearing leads to a contribution of
\beq
\Delta Q_1=\Delta x_1 \frac{\partial Q_1(x_0)}{\partial x_0} =\wvname\frac{\partial Q_1(x_0)}{\partial x_0}\left.\frac{\partial x_1(y)}{\partial y}\right|_{y=y_1}=\frac{\wvname}{\sqrt{M(x_0)}}\frac{\partial Q_1(x_0)}{\partial x_0} \label{poss}
\eeq
to the energy width of the antikink.   This expression is somewhat unappealing for the following reason.  The quantity $x_1(y_1)$ is the centroid of the wave packet in moduli space.  It is an observable quantity.  However $Q_1$ is a function of the arbitrarily chosen modulus $x_0$, which was used to define the kink Hamiltonian $H\p$ via $\mathcal{D}_{\Phi_{x_0}}$.  Thus this expression depends strongly on our choice of $x_0$.   

As we have noted above, the series expansion for the antikink energy converges more rapidly if we choose $x_0$ to be equal to the wave packet centroid $x_1(y_1)$, which implies $y_1=0$.  We will make this choice.  In other words, the kink Hamiltonian $H\p$ is defined by conjugating the defining Hamiltonian $H$ by $\mathcal{D}_{\Phi_{x_0}}$ where $x_0$ is the centroid of the classical modulus of the wave packet considered.  With this convention understood, in the sequel we will often abuse our notation and replace $x_1$ with $x_0$.

%We can now rewrite Eq.~(\ref{x1def}) as follows.  Recall that our moduli space coordinate $x_0$ is defined by
%\beq
%\Phi_{x_0}(x_0)=0.
%\eeq
%This implies that
%\beq
%\left.\frac{\partial \Phi_{x_0}(x)}{\partial x_0}\right|_{x=x_0}
%=-\left.\frac{\partial \Phi_{x_0}(x)}{\partial x}\right|_{x=x_0}.
%\eeq
%Inserting this identity into (\ref{x1def}) one arrives at
%\beq
%\left|\frac{\partial x_0(y)}{\partial y}\right|=\left|\frac{g_B(x_0)} {\left.\partial \Phi_{x_0}(x)/\partial x\right|_{x=x_0}}\right|. \label{repsimp}
%\eeq

Next, consider the smeared momenta. The states $|\wvname\rangle$ are not eigenstates of the $\pi_0^2/2$ term in $H_2$.  To evaluate the corresponding contribution to the energy, one may Fourier transform into a basis of $\pi_0$ eigenstates.  Now the coefficient is a Gaussian with standard deviation $1/\wvname$.  The expectation value of $\pi_0^2/2$ is just half of the variance, and so contributes
\beq
\Delta Q=1/(2\wvname^2)
\eeq
to the expectation value of the energy.  The variance of the energy is then $1/(2\wvname^4)$ and so the momentum smearing contributes $1/\sqrt{2}\wvname^2$ to the energy width.

In all, the width of the energy is
\beq
\Sigma=\sqrt{\frac{\wvname^2}{M(x_0)}\left(\frac{\partial Q_1(x_0)}{\partial x_0}\right)^2+\frac{1}{2\wvname^4}}
\eeq
which simplifies, far from an impurity, to
\beq
\Sigma=\sqrt{\frac{\wvname^2}{Q_0}\left(\frac{\partial Q_1(x_0)}{\partial x_0}\right)^2+\frac{1}{2\wvname^4}}.
\eeq
Recall that, up to factors of order unity, and letting $\alpha\sim 1$ and $l\sim 1/m$
\beq
Q_0\sim \frac{m}{\lambda}\hsp 
Q_1\sim m\hsp
\frac{\partial Q_1(x_0)}{\partial x_0}\sim m^2.
\eeq
Therefore the energy width is minimized for a wave packet of width
\beq
\rho\sim \lambda^{-1/6}m^{-1/2}\hsp \Sigma\sim \lambda^{1/3}m
\eeq
For the range of validity of the semiclassical expansion, $\lambda<<1$, this implies that the energy smearing is smaller than the one-loop correction to the antikink mass, but necessarily larger than the two-loop correction $Q_2\sim m\lambda$ when $\alpha\sim 1$.  

Recalling that $x\sim y/\sqrt{Q_0} \sim \sqrt{\lambda/m}y$ far from an impurity, a width of the wave packet of $\rho\sim \lambda^{-1/6}/\sqrt{m}$ in the $y$ coordinate corresponds to a wave packet of coordinate width $\Delta x\sim \lambda^{1/3}/m$.  This is much larger than the kink's de Broglie wavelength $\lambda/m$, but it is fortunately much smaller than its classical size $1/m$.  Thus, in the semiclassical limit $\lambda<<1$, such a wave packet indeed resembles a classical kink.

The function $Q_1(x_0)$ is only a reasonable approximation to the one-loop correction to the energy if $Q_1(x_0)>>\Sigma$, which requires, far from an impurity
\beq
\frac{\sqrt{Q_0}Q_1(x_0)}{(\partial Q_1(x_0)/\partial x_0)}>>\wvname>>\frac{1}{\sqrt{Q_1(x_0)}}.
\eeq
Such an interval only exists if the length scale $Q_1/(\partial Q_1(x_0)/\partial x_0)$, which is determined by $m$ and the width of the impurity, is greater than the reciprocal of the geometric mean of $Q_0$ and $Q_1(x_0)$.  Any fine structure in $Q_1(x_0)$ smaller than this length scale cannot be probed by antikinks of any $\wvname$.  For example, if the impurity is small and $\alpha\sim 1$, then $1/\sqrt{Q_0Q_1}\sim \sqrt{\lambda}/m$ and $Q_1/(\partial Q_1(x_0)/\partial x_0)\sim 1/m$.   At weak coupling $\lambda<<1$ this inequality is indeed satisfied.%At weak coupling this scale is larger, by the inverse square root of the dimensionless coupling, than the naive de Broglie wave length of the antikink.

More generally, if we allow the length scale of the impurity to differ parametrically from that of the kink $l\sim \lambda^{-n}/m$, for some $n$ then
\beq
\frac{\partial Q_1(x_0)}{\partial x_0}\sim \frac{m}{l} \sim m^2\lambda^{n}
\hsp
\rho\sim \lambda^{-n/3-1/6}m^{-1/2}\hsp \Sigma\sim \lambda^{(2n+1)/3}m.
\eeq
Therefore, for a thin or fat impurity with $n=3k/2-2$, the wave packet smearing affects the energy at the same order as the $k$-loop perturbative correction.

\section{Dynamics} \label{dynsez}

%%%%%%%%%%%%%%%%%%%%%%%%%%%%%%%%%%
\subsection{Acceleration} \label{accsez}
%%%%%%%%%%%%%%%%%%%%%%%%%%%%%%%%%%
We have seen that $Q_1$ depends on the position modulus $x_0$, which is intuitively the intercept of the antikink solution at the centroid of the wave packet.  In the case of a BPS antikink, considered in this note, $Q_0$ is independent of $x_0$ but $Q_1$ continues to depend on $x_0$.   This dependence describes the energy of a wave packet whose center is sufficiently localized about some $x_0$ so that $dQ_1/dx_0$ can be ignored in a leading approximation, and yet sufficiently delocalized that to a leading approximation it may be usefully treated as a Hamiltonian eigenstate.  In this regime, we can consider the dynamical evolution of the antikink and will now calculate its instantaneous acceleration.

States in our antikink basis evolve via the action of the evolution operator
\beq
|\wvname(t)\rangle=e^{-iH\p t}|\wvname(0)\rangle.
\eeq
Since we are trying to calculate the instantaneous acceleration of the antikink, we need only consider the second order evolution
\beq
|\wvname(t)\rangle-|\wvname(0)\rangle\supset -\frac{t^2}{2}H^{\prime 2}|\wvname(0)\rangle \, .
\eeq
The kink Hamiltonian consists of many terms which annihilate $|\wvname(0)\rangle$, as well as the following two terms
\beq
	H' \supset \frac{\pi_0^2}{2} + Q_1\left(x_1(y_1)\right) \, .
\eeq
So when we act on the ground state, we need only consider those two terms. We may promote the position of our wave packet $y_1$ to the operator $\phi_0$. This is allowed since, up to corrections coming from the width of the wave packet, the eigenvalue of $\phi_0$ is $y_1$ and so for all functions $F(\phi_0)| y_1 \rangle = F(y_1) | y_1 \rangle$ holds. In a multiloop calculation, we expect that the one-loop correction $Q_1(y_1)$ will be promoted to an operator $Q_1(\phi_0)$. In that case, this approximation will become exact. Using this approximation, and only keeping terms which will lead to a change in the instantaneous acceleration, we find that
\begin{align} \label{eq:pi0Q1}
		\begin{aligned}
|\wvname(t)\rangle-|\wvname(0)\rangle &\supset -\frac{t^2}{2}\left( \frac{\pi_0^2}{2} + Q_1(\phi_0) \right)^2|\wvname(0)\rangle   \\
&\supset  -\frac{t^2}{2}\left( \frac{\pi_0^2}{2}Q_1\left(\phi_0\right) + Q_1\left(\phi_0\right) \frac{\pi_0^2}{2} \right) |\wvname(0)\rangle .
\end{aligned}
\end{align}
Intuitively, only these terms contribute to the acceleration since the force on the antikink must come from translations of the antikink. Since the wave packet is concentrated around $y_1$, we can approximate \eqref{eq:pi0Q1} by taking a Taylor Series
\beq \label{eq:pi0Q1Taylor}
|\wvname(t)\rangle-|\wvname(0)\rangle\supset -\frac{t^2}{4} \left( Q_1(y_1)\pi_0^2 + \frac{1}{4} \frac{\partial Q_1}{\partial y_1} \{ \phi_0, \pi_0^2 \} \right) |\wvname(0)\rangle \, .
\eeq
Once again, the first term in \eqref{eq:pi0Q1Taylor} doesn't contribute to the acceleration and can be ignored, giving
\begin{align}
		\begin{aligned}
|\wvname(t)\rangle-|\wvname(0)\rangle&\supset    -\frac{t^2}{4\wvname \sqrt{\pi}} \frac{\partial Q_1}{\partial y_1} \{ \phi_0, \pi_0^2 \}   \int dy \exp{-\frac{(y-y_1)^2}{2\wvname^2}}|y\rangle_0 \\
&= -\frac{1}{\wvname \sqrt{\pi}}\frac{\partial Q_1}{\partial y_1}\int dy \frac{t^2 y}{2\wvname^2} \exp{-\frac{(y-y_1)^2}{2\wvname^2}}|y\rangle_0  + O(\hbar) .
	\end{aligned}
\end{align}
We have used the commutation relations \eqref{eq:commutation} to simplify the anti-commutator to $\{ \pi_0^2, \phi_0 \} = 2 \pi_0^2 \phi_0 + O(\hbar)$. We can then move $|\wvname(0)\rangle$ to the right hand side to find
\begin{align}
	\begin{aligned}
|\wvname(t)\rangle &= \frac{1}{\wvname\sqrt{\pi}}\int \left( 1 - \frac{\partial Q_1}{\partial y_1}\frac{t^2y}{2\wvname^2}\right) \exp{ -\frac{(y-y_1)^2}{2\wvname^2} } | y \rangle_0
\\ &\approx \frac{1}{\wvname\sqrt{\pi}}\int \left( 1 - \frac{\partial Q_1}{\partial y_1} \frac{t^2y}{2\wvname^2} - \frac{(y-y_1)^2}{2\wvname^2} \right)  | y \rangle_0 \\ &\approx \frac{1}{\wvname\sqrt{\pi}} \int \exp{ -\frac{(y-y_t)^2}{2\wvname^2} } \, ,
\end{aligned}
\end{align}
where we have expanded and contracted the exponential function. Completing the square, one finds that
\beq
y_t=y_1 - \frac{\partial Q_1}{\partial y_1} \frac{ t^2}{2}.
\eeq
We are thus tempted to identify $-Q_1\p(y_1)$ with the instantaneous acceleration of the antikink.  Actually this is the second time derivative not of the center of mass $x_1(t)$ of the antikink, but rather of the eigenvalue $y_1$ of $\phi_0$.  Recall from Eq.~(\ref{x1def}) that $x_1(y_1)$ and $y_1$ are related by a factor of $1/\sqrt{M(x_0)}$ and so the acceleration is $-Q_1\p(x_0)/M(x_0)$.

To understand the physics behind this result, let us consider an antikink which is well separated from the impurity.  Then, by Eqs.~(\ref{x1def}) and (\ref{q0m}), the derivative of the wave packet centroid $x_1(y_1)$ with respect to $y_1$ is $1/\sqrt{Q_0}$, and so the acceleration is $-Q_1'(y_1)/\sqrt{Q_0}$.   Similarly
\beq
\frac{\partial Q_1(x_1(y_1))}{\partial y_1}=\frac{\partial Q_1(x_1(y_1))}{\partial x_1}\frac{\partial x_1(y_1))}{\partial y_1}=\frac{1}{\sqrt{Q_0}}\frac{\partial Q_1(x_1(y_1))}{\partial x_1}.
\eeq
Combining these results
%The function $Q_1(y_1)$ may be rewritten as a function of the center of mass $x$ of the kink.  This same factor of $\sqrt{Q_0}$ relates the $y_1$ and $x$ derivatives of $Q_1$.   Combining these two factors of $\sqrt{Q_0}$, and using (\ref{cdef}), we find that the instantaneous acceleration is
\beq
\frac{d^2 x(t)}{dt^2}=-\frac{1}{\sqrt{Q_0}}\frac{\partial Q_1(x_1(y_1))}{\partial y_1}=
-\frac{1}{Q_0}\frac{\partial Q_1(x)}{\partial x}.
\eeq
We have recovered Newton's second law $a=F/m$, where $Q_0$ is the leading contribution to the mass and $-Q_1\p(x)$ is the force.  Note that close to the impurity this formula breaks down as $Q_0$ is no longer a good approximation to the effective mass $M(x_0)$ of the antikink.  Evidently, in the presence of an impurity the inertial mass should be identified with $M(x_0)$.  As further evidence for this, we note that the identification of the kinetic term $\pi^2_0/2$ in $H\p_2$ with the nonrelativistic $p^2/2M$ implies that the $\phi_0$-eigenvalue $y_1$ is related to the position $x_1$ by a factor of $\sqrt{M}$.

Overall, we see that the repulsive potential $Q_1(x_0)$ leads to a repulsive force which pushes the antikink away from the impurity. The dramatic classical effects of the spectral wall are seen in models without forces. Hence just the existence of this repulsion suggests that spectral wall will be less pronounced once quantum corrections are included. Understanding the link between quantum corrections and classical forces is difficult, and we have used many approximations in this section. However, our formalism does provide the first steps needed to begin a more careful treatment of these ideas.

\subsection{Exciting the Shape Mode} \label{classsez}

So far we have studied the antikink ground state.  We found that the impurity model considered here, which hosts a spectral wall for sufficiently large $\alpha$, yields an $x_0$-dependent one-loop contribution to the energy of an antikink in its ground state.  This causes a ground state antikink to be repelled from such impurities.  On the other hand, the classical energy $Q_0$ of a BPS antikink is independent of the modulus $x_0$ and so classically there is no such repulsion.  The spectral wall phenomena arises from the observation that, when the shape mode is excited, a classical antikink is repelled from an impurity. If the shape mode is not excited, the antikink can pass through the wall.  In the present section we will try to understand how exciting the shape mode affects the dynamics of the antikink.

Let us reconsider our model
\beq
L=\int_{-\infty}^\infty \left( \frac{1}{2} (\partial_\mu \phi)^2 - \frac{m^2}{4\lambda} \left(\frac{ \left(1-\lambda \phi^2 \right)}{\sqrt{2}} + \alpha  \sech^2(lx)\right)^2 +  \frac{m}{ \sqrt{2\lambda}} \phi \partial_x \alpha  \sech^2(lx) \right) dx, \label{Lag}
\eeq
again with the choice $x_0=x_1$ so that  $y_1=0$.  We will start by introducing the quantum state corresponding to the classical antikink with an excited shape mode.

%Let us recall how the classical kink solution is encoded in the Fourier transformed form factor of a quantum kink.  The kink ground state is, at leading order, $|K\rangle=\df \vac_0$, where $\vac_0$ is annihilated by $B_k$ and $B_S$.  Therefore
%\beq
%\langle K|\phi(x)|K\rangle={}_0\langle 0|\df^\dag \phi(x)\df\vac_0=\langle K|K\rangle f(x)+g_B(x){}_0\langle 0|\phi_0\vac_0.
%\eeq
%Let us replace $\vac_0$ by a normalized wave packet $|y\rangle$ satisfying
%\beq
%\langle y |y\rangle=1\hsp 
%\langle y |\phi_0|y\rangle=y.
%\eeq
%The state $|y\rangle$ is approximately equal to $\vac_0$ in the sense that it can be written as a power series in $\phi_0$ acting on $\vac_0$ with the constant term equal to unity, or more concretely it is an integral over rapidities $\eta$ of boosted states $\Lambda_\eta\vac_0$ where $\Lambda_\eta$ is a boost operator.  Define
%\beq
%|K\rangle_y=\df|y\rangle.
%\eeq
%Then
%\beq
%{}_y\langle K|\phi(x)|K\rangle_y=f(x)+y g_B(x)
%\eeq
%up to corrections arising from the fact that the wave packet is not the true ground state, which were estimated above.  Note that this leading contribution only dominates for sufficiently small $y$, where the second term can be absorbed into the first by using the classical solution at a different value of the modulus.  For simplicity, let us choose the value of the modulus at which f(x) is defined so that $y=0$.

The operator $B^\dag_S$ excites the shape mode.  Let us define the coherent state $|\beta\rangle$ to be the normalized state
\beq
|\beta\rangle=\rm{exp}\left[\beta \left(\sqrt{2\omega_S}B^\dag_S- \frac{B_S}{\sqrt{2\omega_S}}\right)\right]|y\rangle_0\hsp
|K\rangle_\beta=\df|\beta\rangle .
\eeq
Using\footnote{Recall our convention that the adjoint of $B_S$ is $2\omega_S B^\dag_S$.}
\beq
B_S|\beta\rangle=\sqrt{2\omega_S}\beta|\beta\rangle\hsp \langle\beta|B_S^\dag=\langle\beta|\frac{\beta}{\sqrt{2\omega_S}}\hsp
 \langle\beta|\phi(x)|\beta\rangle=\beta\sqrt{\frac{2}{\omega_S}}g_S(x),
\eeq
one finds
\beq
{}_\beta\langle K|\phi(x)|K\rangle_\beta=\Phi_{x_0}(x)+\beta\sqrt{\frac{2}{\omega_S}}g_S(x). \label{clim}
\eeq
This is our master formula for the normalization of quantum shape modes.

Using the $\omega_SB^\dag_S B_S$ term in $H\p_2$, the one-loop expectation value of the energy of the shape mode is
\beq
\langle \beta|\omega_SB^\dag_S B_S|\beta\rangle
=\beta^2\omega_S.
\eeq
Substituting the second term in (\ref{clim}), interpreted as the classical field, into the Hamiltonian of the classical field theory and using the linearized classical wave equation satisfied by $g_S(x)$, one finds that the classical energy is $\beta^2\omega_S$.

How does this compare to the moduli space approach of Ref.~\cite{muri}?  As the shape mode $g_S$ here is normalized identically to the collective coordinate $\eta$ there, one may identify $\beta\sqrt{2/\omega_S}$ with the amplitude of oscillation of the collective coordinate $A(t)$.
%\beq
%A(t)=\frac{\beta\sqrt{2}}{\sqrt{\omega_S}}.
%\eeq
In  Eq.~(5) of Ref.~\cite{muri} it was found that in the moduli space approximation, the system is described by the Hamiltonian
\beq
H=\frac{1}{2}\left[\dot{A}^2(t)+M_{x_0}(t)\dot{x}_0^2(t)+\omega_{S,x_0}^2 A^2(t)
\right]\hsp M_{x_0}(t)=\int dx \left[\partial_{x_1}\left(f_{x_1}(x)+A(t)g_{S,x_1}(x)\right)|_{x_1=x_0}\right]^2 \label{collh}
\eeq
where we have made the dependence of $g_S(x)$ and $\omega_S$ on the modulus $x_0$ explicit.  In the case of a stationary antikink with constant shape mode amplitude, the energy contribution of the shape mode to a stationary antikink is just equal to the first plus last term
\beq
H=\frac{1}{2}\left[\dot{A}^2(t)+\omega_S^2 A^2(t)\right]=\frac{1}{2}\left[\omega_S^2 \left(\frac{\beta\sqrt{2}}{\sqrt{\omega_S}}
\right)^2
\right]=\beta^2\omega_S.
\eeq
Thus, all three expressions for the classical energy agree. 

\begin{figure}[h] %[!tph]
	\begin{center}
\includegraphics[width=3in]{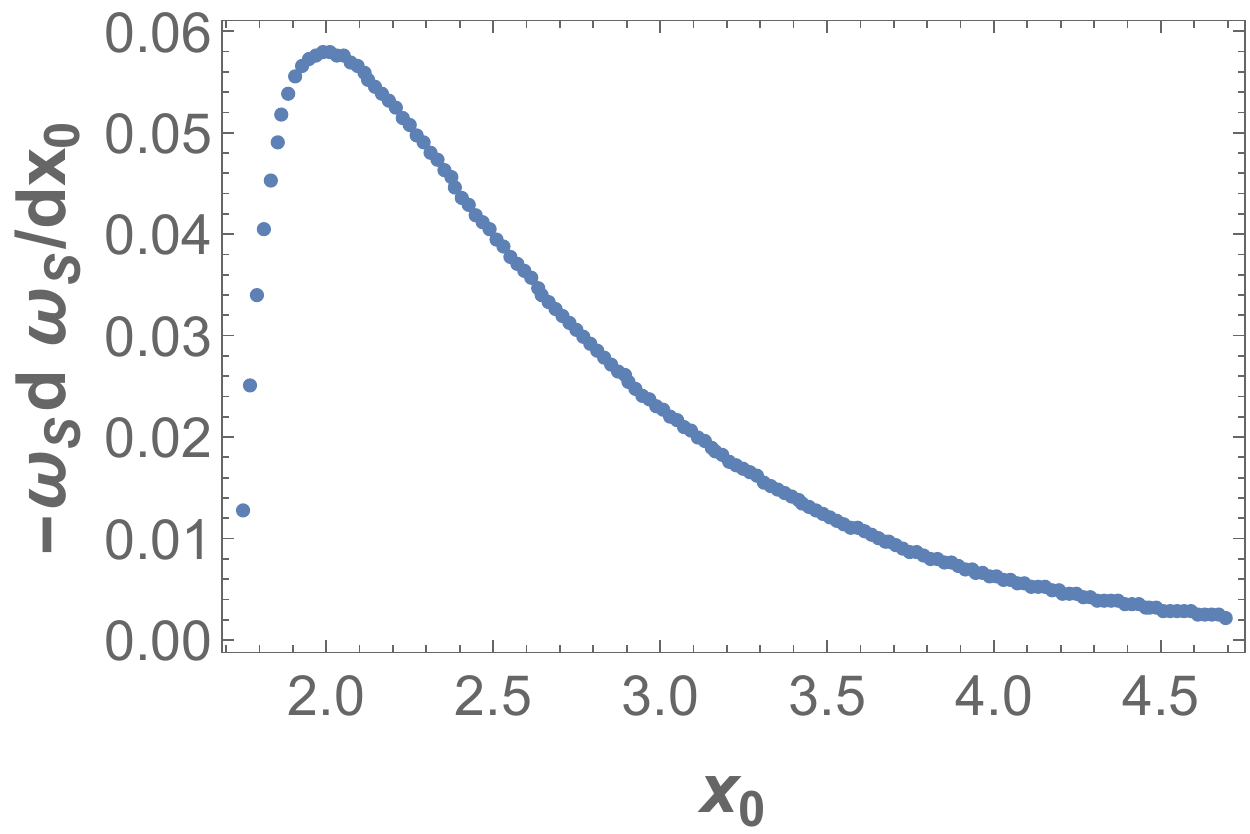}
		\caption{The force constant $-\omega_S \frac{d\omega_S}{dx_0}$ at $\alpha=3$, $m=2$ and $l = \lambda=1$, which determines the force exerted on a quantum antikink whose shape mode is described by a coherent state, or equivalently on a classical antikink with an excited shape mode.  This describes the contribution to the force resulting from the excited shape mode and does not include the contributions from zero-point energies which were described in previous sections.}
		\label{wdwfig}
	\end{center}
\end{figure}

What about the instantaneous acceleration $\ddot{x}_0(t=0)$?  If the antikink begins at rest, then $\dot{x}_0(t=0)=0$.  Therefore, the $A$ equation of motion from (\ref{collh}) yields
\beq
\ddot{A}(t=0)=-\omega_{S,x_0}^2 A(t=0).
\eeq
Therefore, any change in $x_0$ only affects $A(t)$ at order $O(t^3)$, and so does not contribute to the instantaneous acceleration.  The equation of motion for the modulus $x_0$ derived from (\ref{collh}) is just Newton's Second Law
\beq
M_{x_0}(t)\ddot{x}_0(t)=-\frac{\partial M_{x_0}(t)}{\partial x_0}\frac{\dot{x}_0^2(t)}{2}
-\dot{M}_{x_0}(t)\dot{x}_0(t)- \omega_S\frac{d\omega_S}{dx_0} A^2(t)
\eeq
where the right hand side is the force.  Our initial condition, that the antikink is at rest, means that the first two terms initially vanish, leaving an instantaneous force at time zero of
%As was already noted in Ref.~\cite{muri}, it is $A$, not $\beta$, that has a canonical kinetic term.  Therefore the force is obtained from minus the derivative of the energy with respect to $x_0$ taken holding $A$ constant
\beq
F=-\omega_S \frac{d\omega_S}{dx_0} A^2 .
\eeq
This is plotted numerically in Fig.~\ref{wdwfig}. We see that this is a repulsive force, meaning that the kink is further repelled from the impurity when the shape mode is excited. Note that the force appears to vanish at the wall $x_{sw} = 1.65$, and attains its maximum just outside of the wall.  Inside of the wall there is no shape mode and so no corresponding force.  The leading contribution to the acceleration is the force divided by the classical energy of the excited solution,  which consists of all three mass terms in Eq.~(5) of Ref.~\cite{muri}.  

It may be unexpected that the force is the derivative of $\omega_S^2$ and not $\omega_S$, as it would be were $\beta$ fixed instead of the amplitude of $A$.  As $\beta$ is the dimensionless expected quantum number of the shape mode of the excitation, one might think that during acceleration $\beta$ should be fixed rather than the dimensionful collective coordinate $A$.  The fact that the amplitude of $A$ is fixed up to order $t^2$ implies that as an antikink approaches a spectral wall, and so as $\omega_S$ grows, the quantum number of the shape mode excitation grows too.  Thus we see that the excitation number and the energy cost per excitation both grow linearly in $\omega_S$, resulting in an energy proportional to $\omega_S^2$.  This surprising behavior may result from the fact that, near the wall, the shape mode is very delocalized and any classical profile at fixed mode number is similarly delocalized.  Thus if one fixes a classical profile for the shape mode, then as it approaches the wall its mode decomposition requires higher and higher modes.

\section{Conclusion and Future Directions}

We have studied a single, stationary quantum antikink in the presence of an impurity.  We have found that, for the values of the impurity strength $\alpha$ considered and perhaps more generally for $\alpha>0$, the antikink always is repelled by the impurity.  This repulsion exists whether or not the shape mode is excited, although it is stronger when the shape mode is excited.  

%We have found that nothing unexpected happens at the spectral wall point, $x_{sw}$. 
We have found that nothing unexpected happens at the spectral wall, $x_{sw}$. The classical dynamics is greatly affected by the spectral wall and the Hamiltonian \eqref{h2p} naively looks like it changes discontinuously across the wall. However, we found that the one-loop correction is continuous across $x_{sw}$. A possible explanation for this fact is that the quantum effect induces a force on the moduli space, breaking the BPS feature of the model. Classically, when the model is not BPS the spectral walls become \emph{thick} walls \cite{Thick} which exist across a finite interval of space and whose position depends on the velocity of the incoming antikink. These features naturally smooth out any discontinuous behaviour at the spectral wall.

Our calculation was performed using the general framework of kink sector perturbation theory in Refs.~\cite{mekink}.  This framework still has some quite serious limitations.  For example, it is only reliable in the vicinity of some chosen base point in moduli space.  This base point can be chosen arbitrarily, but once it is chosen it is fixed.  Therefore it is not yet possible to follow a moving kink, or to consider the dynamics of multiple kinks.  This is the reason that only instantaneous acceleration can be computed.  That said, if one knows the instantaneous acceleration for each kink position and each kink momentum, then the acceleration can be integrated to yield a trajectory and so in principle it would be possible to study kink-impurity scattering even in the quantum theory.  

It would also be possible to follow an excited quantum antikink as it passes a wall, observing how the shape modes transform into continuum modes.  Using the explicit state and Hamiltonian constructed here, such an evolution is quite straightforward.  One merely applies $e^{-iHt}$ to the initial state.  The interplay between the momentum and zero mode and also the shape and continuum modes may be simplified by considering a stationary antikink and a quench in which the shape mode dissolves into the continuum.  This is quite similarly straightforward with Hamiltonian evolution.  The study of kinks in quenched theories has recently received attention in Ref.~\cite{horvath21}.

Beyond one loop, using the formalism of Ref.~\cite{me2loop}, the different normal modes couple to one another.  One can therefore expect a much richer phenomenology in these systems, considering that the momentum of moving kinks can also be transferred into their normal modes.  Higher excitations of shape modes also decay into continuum modes beyond one loop, which will render the dynamics of excited quantum kinks yet more interesting.

The focus of this method on time-independent Hamiltonian eigenstates also limits the systems to which it may be applied at present.  While this manuscript was in preparation, Ref.~\cite{kovtun21} appeared which studied a one-loop time-dependent nontopological soliton, building upon the studies of such normal modes in Ref.~\cite{kovtun18}.  A systematic approach to higher-loop corrections in such cases would be interesting to pursue.

\section* {Acknowledgement}

\noindent
JE is supported by NSFC MianShang grants 11875296 and 11675223 and the CAS Key Research Program of Frontier Sciences grant QYZDY-SSW-SLH006.   CH is supported by The University of Leeds as an Academic Development Fellow. TR and AW were supported by the Polish National Science Centre, 
grant NCN 2019/35/B/ST2/00059.

%%%%%%%%%%%%%%%%%%%%%%%%%%%%%%%%%%
\appendix
%%%%%%%%%%%%%%%%%%%%%%%%%%%%%%%%%%
\section{From Plane-Wave to Normal-Mode Normal-Ordering} \label{appn}
\subsubsection{Zero Modes}

We will convert $H_2$, defined in \eqref{eq:H2den}, from plane-wave to normal-mode normal-ordering, one term at a time.   First consider the $\pi^2$ term
\beq
\int dx\frac{:\pi^2(x):_a}{2}=\pin{p} \frac{:\pi_p\pi_{-p}:_a}{2}=\pin{p}\frac{\op^2}{2}\left[- A^\dag_{-p}\left(A^\dag_{p}-\frac{A_{-p}}{2\op}\right)
+\left(A^\dag_{p}-\frac{A_{-p}}{2\op}\right)\frac{A_{p}}{2\op}
\right]. \label{pp}
\eeq
The zero mode contribution is
\beq
\int dx \frac{:\pi^2(x):_a}{2} \supset \pin{p}\frac{\tg_B(p)\tg_B(-p)}{4}\left(2\pi_0^2-i\op[\pi_0,\phi_0]\right)=\frac{\pi_0^2}{2}- \pin{p}\left|\tg_B(p)\right|^2\left(\frac{\op}{4}\right) \label{zeroc}
\eeq
where we used
\beq
\pin{p}\tg_B(p)\tg_B(-p)=\pin{p}\int dx\int dy e^{ip(x-y)}g_B(x)g_B(y)=\int dx g^2_B(x)=1
\eeq
and
\beq
\tg_B^*(p)=\tg_B(-p).
\eeq

Next consider the term
\bea
&&\int dx\v2\frac{:\phi^2(x):_a}{2}\\
&&=\int dx\pin{p}\pin{q}e^{-ix(p+q)}\frac{\v2}{2}\left[A^\dag_p\left(A^\dag_q+\frac{A_{-q}}{2\oq}\right)+\left(A^\dag_q+\frac{A_{-q}}{2\oq}\right)\frac{A_{-p}}{2\op}\nonumber
\right].
\eea
The zero mode contribution is
\bea
&&\int dx\v2\frac{:\phi^2(x):_a}{2}\label{zero}\\
&&\supset\int dx\pin{p}\pin{q}e^{-ix(p+q)}\frac{\v2}{4}\tg_B(p)\tg_B(q)\left[
\phi_0^2-\frac{i}{\op}[\pi_0,\phi_0]
\right]\nonumber\\
&&=\int dx\pin{p}e^{-ixp}\frac{\v2 g_B(x)}{4}\tg_B(p)\left[
\phi_0^2-\frac{1}{\op}\right]\nonumber\\
&&=\int dx\pin{p}e^{-ixp}\frac{\partial_x^2g_B(x)}{4}\tg_B(p)\left[
\phi_0^2-\frac{1}{\op}\right].\nonumber
\eea

The last term in $H\p_2$ is
\bea
&&\int dx\frac{:(\partial_x\phi(x))^2:_a}{2}\\
&&=\int dx\pin{p}\pin{q}e^{-ix(p+q)}\frac{q^2}{2}\left[A^\dag_p\left(A^\dag_q+\frac{A_{-q}}{2\oq}\right)+\left(A^\dag_q+\frac{A_{-q}}{2\oq}\right)\frac{A_{-p}}{2\op}\nonumber
\right]
\eea
whose zero mode part  is
\bea
&&\int dx\frac{:(\partial_x\phi(x))^2:_a}{2}\supset \int dx\pin{p}\pin{q}e^{-ix(p+q)}\frac{q^2}{4}\tg_B(p)\tg_B(q)\left[
\phi_0^2-\frac{1}{\op}
\right]
\eea
which exactly cancels the zero mode contribution in Eq.~(\ref{zero}).

Therefore the zero mode contribution to $H_2$ is given by Eq.~(\ref{zeroc}).  Note, if the connected component of the moduli space under consideration consists of isolated points, as we will see below in a vacuum sector, then there is no zero mode and so there is no such contribution.

\subsubsection{Shape Modes}

Next we will calculate the shape mode contributions to these three terms.   First, inserting the Bogoliubov transformation into Eq.~(\ref{pp}), we find
\bea
\int dx \frac{:\pi^2(x):_a}{2} &\supset& \pin{p}\frac{\tg_S(p)\tg_S(-p)}{2} \label{pshape}\\
%&&\times\left[-\frac{\os}{\op}B_S^{\dag 2}
%+\left(1+\frac{\os}{\op}\right)B^\dag_S\frac{B_S}{2\os}
%+\left(-1+\frac{\os}{\op}\right)\frac{B_S}{2\os}B^\dag_S
%-\frac{\os}{\op}\left(\frac{B_S}{2\os}\right)^2
%\right]\nonumber\\
&&\times
\left[-\os^2 B_S^{\dag 2}
+2\os^2B^\dag_S\frac{B_S}{2\os}
-\os^2\left(\frac{B_S}{2\os}\right)^2
+\left(\frac{\os}{2}-\frac{\op}{2}\right)
\right].\nonumber
%= \pin{p}\tg_B(p)\tg_B(-p)\left(\frac{\pi_0^2}{2}-\frac{\op}{4}\right).
\eea
Next, the shape mode contribution to the potential term is
\bea
&&\int dx\v2\frac{:\phi^2(x):_a}{2}\\
&&\supset\int dx\pin{p}e^{-ixp}\frac{\v2 g_S(x)}{2}\tg_S(p)%\nonumber\\
%&&\times\left[B_S^{\dag 2}+\left(1+\frac{\os}{\op}\right)B^\dag_S\frac{B_S}{2\os}+\left(1-\frac{\os}{\op}\right)\frac{B_S}{2\os}B^\dag_S+\left(\frac{B_S}{2\os}\right)^2\right]\nonumber\\
%&&\times
\left[
B_S^{\dag 2}
+2B^\dag_S\frac{B_S}{2\os}
+\left(\frac{B_S}{2\os}\right)^2
+\left(\frac{1}{2\os}-\frac{1}{2\op}\right)
\right]\nonumber\\
&&=\int dx\pin{p}e^{-ixp}\frac{(\os^2+\partial_x^2)g_S(x)}{2}\tg_S(p)
%\nonumber\\&&\times
\left[
B_S^{\dag 2}
+2B^\dag_S\frac{B_S}{2\os}
+\left(\frac{B_S}{2\os}\right)^2
+\left(\frac{1}{2\os}-\frac{1}{2\op}\right)
\right].\nonumber
\eea
Again the $\partial_x^2$ term cancels the contribution from $(\partial\phi)^2/2$, leaving
\beq
\pin{p}\frac{\tg_S(-p)\tg_S(p)}{2}
%\nonumber\\&&\times
\left[
\os^2B_S^{\dag 2}
+2\os^2B^\dag_S\frac{B_S}{2\os}
+\os^2\left(\frac{B_S}{2\os}\right)^2
+\left(\frac{\os}{2}-\frac{\os^2}{2\op}\right)
\right].
\eeq
The first and third terms cancel those in (\ref{pshape}), leaving a shape mode contribution to the kink Hamiltonian of
\beq
\os B^\dag_SB_S
- \pin{p}\left|\tg_S(p)\right|^2
\left[
\frac{(\os-\op)^2}{4\op}
\right].
\eeq
Again, it is often the case that there are no shape modes, in which case there is no such contribution.

\subsubsection{Continuum Modes}
The calculation for the continuum modes is very similar, although now the $g_k(x)$ are complex as each describes two normal modes.  The continuum contribution to the $\pi^2$ term is
\bea
\int dx \frac{:\pi^2(x):_a}{2} &\supset& %\pin{p}\pink{2}\frac{\tg_{k_1}(p)\tg_{k_2}(-p)}{4}\ok{1} \label{pshape}\\
%&\times&\left[-2\ok{2}\Bd{1}\Bd{2}
%+\left(\op+\ok{2}\right)\left(\Bd1\frac{B_{-k_2}}{2\ok2}+\Bd2\frac{B_{-k_1}}{2\ok1}\right)\right.\nonumber\\
%&&\left.
%+\left(-\op+\ok{2}\right)\left(\frac{B_{-k_2}}{2\ok2}\Bd1+\frac{B_{-k_1}}{2\ok1}\Bd2\right)
%-2\ok{2}\frac{B_{-k_1}B_{-k_2}}{4\ok{1}\ok{2}}
%\right]\nonumber\\
%&=& 
\pin{p}\pink{2}\frac{\tg_{k_1}(p)\tg_{k_2}(-p)}{2}\ok{1} \\
&\times&\left[-\ok{2}\Bd{1}\Bd{2}
+\ok{2}\left(\Bd1\frac{B_{-k_2}}{2\ok2}+\Bd2\frac{B_{-k_1}}{2\ok1}\right)\right.\nonumber\\
&&\left.
-\ok{2}\frac{B_{-k_1}B_{-k_2}}{4\ok{1}\ok{2}}
+\left(-\op+\ok{2}\right)\left(\frac{2\pi\delta(k_1+k_2)}{2\ok2}\right)
\right].\nonumber
\eea
Using
\beq
\pin{p}\tg_{k_1}(p)\tg_{k_2}(-p)=\int dx\int dy \pin{p} e^{ip(x-y)} g_{k_1}(x)g_{k_2}(y)=2\pi\delta(k_1+k_2)
\eeq
the $\pi^2$ contribution simplifies to
\beq
\pin{k}\frac{\ok{}^2}{2}\left[-B^\dag_kB^\dag_{-k}+2B^\dag_k\frac{B_{k}}{2\ok{}}-\frac{B_kB_{-k}}{4\ok{}^2}\right]
+\pin{p}\pink{2}\tg_{k}(p)\tg_{-k}(-p)\left(\frac{\ok{}}{4}-\frac{\op}{4}
\right).
\eeq
Again the contribution from the kinetic and potential terms cancel the first and third terms, double the second and modify the last, leaving
\beq
\pin{k}\left[\ok{} B^\dag_kB_k
- \pin{p}\left|\tg_k(p)\right|^2
\frac{(\ok{}-\op)^2}{4\op}
\right].
\eeq

%%%%%%%%%%%%%%%%%%%%%%%%%%%%%%%%%%
\section{Numerical Techniques}
\label{Num}
%%%%%%%%%%%%%%%%%%%%%%%%%%%%%%%%%%

All numerics in this paper are done using the Lagrangian \eqref{Lag} with parameters $\lambda=l=1$ and $m=2$.

%%%%%%%%%%%%%%%%%%%%%%%%%%%%%%%%%%
\subsection*{Gradient flow on a finite interval}
%%%%%%%%%%%%%%%%%%%%%%%%%%%%%%%%%%

This method is based on \cite{bt}. It is useful as it can be applied to solitons in dimension greater than one \cite{gh}. We put the system on a finite box of length $L$ and equally discretize space with a lattice spacing $\Delta x$. Instead of solving \eqref{SLfluc} directly, we take a random initial perturbation $\epsilon_0$ and evolve it using
\begin{equation}
	\dot \epsilon = -\left( -\frac{d^2}{dx^2} + V^{(2)}(\Phi_{x_0}(x),x) \right) \epsilon \ . 
\end{equation}
Writing the initial perturbation as a linear combination of eigenvectors $\{ g_i \} = \{ g_B, g_S, g_k \} $ of the operator, its evolution with time becomes obvious:
\begin{equation}
\epsilon_0 = \sum_{i} a_i g_i \implies \epsilon(t) = \sum_{i} a_i g_i e^{-\omega_i t}
\end{equation}
where $\omega_i$ are the eigenvalues of the eigenvectors. Hence, after a long time the lowest frequency mode dominates the dynamics. At this point we extract this mode and repeat the process while projecting out the extracted mode. After extracting the second, we repeat while projecting out the first two modes. And so on. This builds a library of $N$ perturbations ordered by frequency. Since the box has a finite length, there is a discrete set of modes, rather than a continuum as there is on $\mathbb{R}$. On the same grid, we build a library of $N'$ plane waves with frequency $\omega_p$. As this is computationally simple, we take $N'$ very large. 

\begin{figure} %[!tph]
	\begin{center}
		\includegraphics[width=4in]{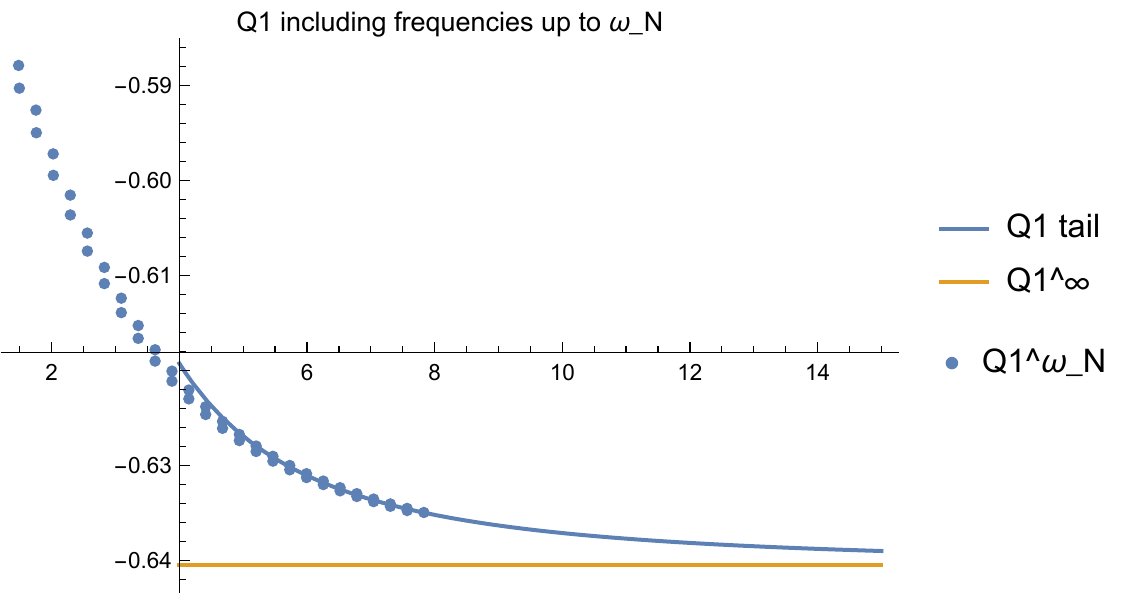}
		\caption{Fitting an analytical tail to our numerical approximation of the one-loop correction as a function of the maximum frequency included. This fit is for $\alpha=0.3$ and $\phi_0=0$.}
		\label{fig:Q1_N}
	\end{center}
\end{figure}

To calculate the one-loop correction \eqref{q1} we discretize the integrals
\begin{equation}
	\int dp \to \sum^{N'}\, , \qquad \int dk \to \sum^N \, .
\end{equation}
Denote the one-loop correction including $N$ modes as $Q_1^{(N)}$. This gives us the one-loop correction up to frequencies of $\omega_N$. The large frequency contribution can be found using a WKB approximation. This shows that
\begin{equation}
	Q_1^{(N)} \sim \frac{1}{\omega_N^2}
\end{equation}
and we fit this tail to our numerically generated density $Q_1^{(n)}$. This density, and a fitted tail is displayed in Figure \ref{fig:Q1_N}. The calculations in this paper are done on a box with grid length $24$ and a lattice spacing of $0.04$. A total $60$ modes are generated for each classical configuration whose mass is calculated.

%%%%%%%%%%%%%%%%%%%%%%%%%%%%%%%%%%
\subsection*{Uncompactified Approach}
%%%%%%%%%%%%%%%%%%%%%%%%%%%%%%%%%%

We also used an approach in which no compactification was imposed.  At each value of $k>0$, there is a two dimensional space of continuum modes.  A grid of values of $k$ was selected.  For each, we began at a point with two distinct sets of real initial conditions for the field and its first spatial derivative.  Then the linearized equation of motion was numerically integrated to generate the solution out to the region where the potential is flat, where the contribution of the kink and impurity are both negligible.  In this region, each solution is a sine wave.  

We considered two such solutions and determined the amplitude of the sine waves at large $x$ and $-x$ for each solution.  Then we constructed two linearly independent, linear combinations of the two solutions which each have the same amplitude at large $x$ and $-x$.  Finally we constructed two linear combinations of those solutions which are orthonormal with respect to integration.  More precisely, the amplitude is unity at large $x$ and $-x$ for each function, and the integral of the product over a sufficiently large interval of positive $x$ values is negative that same integral over negative $x$ values.  The function $g_k(x)$ is defined to have its real and imaginary parts equal to those two functions.

The Fourier transform is performed by first multiplying $g_k(x)$ by a small exponential damping factor, and taking the limit that this damping factor goes to zero.  This limit is taken analytically, as the high $|x|$ region is just a sine wave.

The bound normal modes are treated differently.  For a trial value of the frequency, the eigenvalue equation for normal modes is discretized and written as a matrix equation, which is solved by inversion.  The result does not satisfy the boundary conditions that $g(x)$ should tend to zero at infinity.  To improve  the situation, a new value of the frequency is guessed, even in the case of the zero-mode, and this is repeated a fixed number of times, usually a few hundred or thousand until  the boundary conditions are satisfied out to large $|x|$.

\section{The Vacuum Sector} \label{vacsez}

Here we present an approximated analytical computation of the one-loop energy in the vacuum sector when the impurity is small. Consider a vacuum sector of a model with an impurity.   For a fixed impurity and sufficiently large $k$, the amplitude of the normal modes will be approximately constant.   We will now derive a simple formula for the contribution to $Q_1$ arising at such large $k$, or equivalently at fixed $k$ for a small impurity.

Let the interval $[0,x_f]$ be the spatial support of $\v2$.  In this approximation, the impurity is reflectionless and each normal mode may be divided into three parts
\beq
g_k(x)=\left\{
\begin{tabular}{cc}$e^{-ikx}$ & if\ \ $x<0$\\
	$e^{-i\int dx k\p(k)}$&if\ \ $0<x<x_f$\\
	$e^{-i\left(\delta(k)+kx\right)}$ & if\ \ $x>x_f$
\end{tabular}
\right.
\eeq
where
\beq
\delta(k)=\int_0^{x_f} dx (k\p(k)-k).
\eeq
An analogous expression may be written for the $e^{ikx}$ mode.

For now, let us make the crude approximation that $\v2$ is equal to $V$ in the middle interval and $m^2$ elsewhere.  Then the linearized equations of motion (\ref{cleq}) obeyed by the normal modes yield
\beq
k\p=\sqrt{k^2+m^2-V}\sim k+\frac{m^2-V}{2k} \label{kpeq}
\eeq
where we have used our approximation $k^2>|m^2-V|$.  Thus we find
\beq
\delta(k)=x_f\frac{m^2-V}{2k}.
\eeq

We calculate $\tg_k(p)$ one region at a time.  The first region contributes
\beq
\int_{-\infty}^0 dx g_k(x) e^{ipx}=\left.\frac{i}{k-p} e^{-i(k-p)x}\right|_{-\infty}^0=\frac{i}{k-p}.
\eeq
In the last equality we have dropped the contribution from $x=-\infty$.  This is because it yields a quantity which rapidly oscillates in $p$ at large $|x|$, and so when folded into any continuous function in $p$ it will vanish by the Riemann-Lebesgue Lemma.  This is not true at $p=k$, leading to a $2\pi\delta(p-k)$ which we ignore here as it does not contribute to $Q_1$ \cite{memassa}.  The last equality may alternatively be derived from the usual prescription of including an exponential damping in the integrand, integrating and then taking the damping factor to zero after the integration.  

Similarly the third region contributes
\beq
\int_{x_f}^\infty dx g_k(x) e^{ipx}=-\frac{i}{k-p}e^{-i\left(\delta(k)+(k-p)x_f\right)}.
\eeq
The second contributes
\beq
\int_{0}^{x_f} dx g_k(x) e^{ipx}=-i\left(\frac{1-e^{-i(k\p(k)-p)x_f}}{k\p-p}\right).
\eeq

Combining these contributions we find
\beq
\tg_k(p)=i\left(1-e^{-i(k\p(k)-p)x_f}\right)\frac{k\p(k)-k}{(k-p)(k\p(k)-p)}. \label{tge}
\eeq
Inserting this and (\ref{kpeq}) into (\ref{q1}) yields
\bea
Q_1&=&-\frac{1}{4}\pin{k}\pin{p}\frac{(\ok{}-\omega_p)^2}{\omega_p}\tg_k(p)\tg_{-k}(-p)=-\pin{k} (k\p(k)-k)^2\frac{k^2}{\ok{}^3}\pin{p}\frac{\sin^2\left(\frac{(k\p(k)-p)x_f}{2}\right)}{(k\p-p)^2}\nonumber\\
&=&-\frac{ x_f}{4}\pin{k} (k\p(k)-k)^2\frac{k^2}{\ok{}^3}=-\frac{ x_f}{16}\left(m^2-V\right)^2\pin{k} \frac{1}{\ok{}^3}=-x_f\frac{ \left(m^2-V\right)^2}{16\pi m^2}
\eea
where, going from the second expression to the third, we have used $k\sim p$ to approximate
\bea
(\ok{}-\omega_p)^2&=&\left(k+\frac{m^2}{2k}-p-\frac{m^2}{2}\frac{1}{k-(k-p)}\right)^2=\left(k-p-\frac{m^2}{2}\frac{k-p}{k^2}\right)^2\nonumber\\&=&(k-p)^2\left(1-\frac{m^2}{2k^2}\right)^2=(k-p)^2\frac{k^2}{\ok{}^2}.
\eea
The fact that this is linear in $x_f$ suggests that we may drop the restriction that $\v2$ be constant
\beq
Q_1=-\int dx\frac{ \left(m^2-\v2\right)^2}{16\pi m^2}. \label{guess}
\eeq

We remind the reader that we have approximated $|k|$ to be large, and so we do not expect the small $k$ contribution to $Q_1$ to be well-approximated by this formula.  That said, in any example one can determine just which values of $k$ are reliable by observing the position-dependence of the amplitude in a numerically evaluated normal mode.


\begin{thebibliography}{99}

% solitons 

\bibitem{SkPerr}
J.~K.~Perring and T.~H.~R.~Skyrme,
``A Model unified field equation,"
Nucl. Phys. \textbf{31} (1962), 550.


\bibitem{MS} N. Manton and P. Sutcliffe,
\textit{``Topological Solitons},"
Cambridge University Press, Cambridge U.K. (2004).

\bibitem{Sh} Y. M. Shnir, ``Topological and Non-Topological Solitons in Scalar Field Theories", Cambridge University Press, Cambridge U.K. (2018).

\bibitem{K} P.~G. Kevrekidis and J. Cuevas-Maraver (eds.), ``A Dynamical Perspective on the $\phi^4$ Model," Nonlinear Systems and Complexity, vol. \textbf{26}, Springer Nature, Cham, 2019.

% \phi^4

\bibitem{Sug} T. Sugiyama,
``Kink-antikink collisions in the two-dimensional $\phi^4$ model,"
Prog. Theor. Phys. \textbf{61}, 1550 (1979).

\bibitem{Mosh} M. Moshir, 
``Soliton-antisoliton scattering and capture in $\lambda\phi^4$ theory, "
Nucl. Phys. \textbf{B185} (1981) 318.

% resonant mechanism

\bibitem{CSW} D.~K. Campbell, J.~F. Schonfeld and C.~A. Wingate,
``Resonance structure in kink-antikink interactions in $\phi^4$ theory,"
Physica \textbf{D9} (1983) 1.

\bibitem{MORW} N. S. Manton, K. Oles, T. Romanczukiewicz and A. Wereszczynski, ``Collective coordinate model of kink-antikink collisions in $\phi^4$ theory", Phys. Rev. Lett. {\bf 127} (2021) 071601 [arXiv:2106.05153]. 

% spectral wall

\bibitem{muri}
C.~Adam, K.~Oles, T.~Romanczukiewicz and A.~Wereszczynski,
``Spectral Walls in Soliton Collisions'',
Phys. Rev. Lett. \textbf{122} (2019) no.24, 241601
doi:10.1103/PhysRevLett.122.241601
[arXiv:1903.12100].

% BPS models

\bibitem{Bo} E.~B. Bogomolny,
``The stability of classical solutions,"
\textit{Sov. J. Nucl. Phys.} \textbf{24} (1976) 449.

\bibitem{JT} A. Jaffe and C. Taubes, ``Vortices and
Monopoles," Boston, Birkh\"auser, 1980.

\bibitem{Ma} N.~S. Manton,
``The force between 't Hooft-Polyakov monopoles,"
\textit{Nucl. Phys.} \textbf{B126} (1977) 525.

% Canonical Moduli Space and Dynamics

\bibitem{NM} N.~S. Manton, ``A remark on the scattering of BPS
monopoles," \textit{Phys. Lett.} \textbf{B110} (1982) 54.

\bibitem{AH} M.~F. Atiyah and N.~J. Hitchin,
``The Geometry and Dynamics of Magnetic Monopoles,"
Princeton University Press, Princeton NJ, 1988.

\bibitem{dhn2}
  R.~F.~Dashen, B.~Hasslacher and A.~Neveu,
  ``Nonperturbative Methods and Extended Hadron Models in Field Theory 2. Two-Dimensional Models and Extended Hadrons,''
  Phys.\ Rev.\ D {\bf 10} (1974) 4130.
 doi:10.1103/PhysRevD.10.4130

%\bibitem{Raja}
%R.~Rajaraman,
%``Some non-perturbative semi-classical methods in quantum field theory (a pedagogical review),"
%Physics Reports,
%{\bf 21} (1975), 227. doi:10.1016/0370-1573(75)90016-2

\bibitem{crystal}
Y. Wada, J. R. Schrieffer,  
``Brownian motion of a domain wall and the diffusion constants,''
 Phys. Rev. B \textbf{18}  (1978) no.8, 3897.

\bibitem{gjs}
J.~L.~Gervais, A.~Jevicki and B.~Sakita,
``Perturbation Expansion Around Extended Particle States in Quantum Field Theory. 1.,''
Phys. Rev. D \textbf{12} (1975), 1038
doi:10.1103/PhysRevD.12.1038


\bibitem{me2loop}
J.~Evslin and H.~Guo,
``Two-Loop Scalar Kinks,''
Phys. Rev. D \textbf{103} (2021) no.12, 125011
doi:10.1103/PhysRevD.103.125011
[arXiv:2012.04912 [hep-th]].

\bibitem{cahill76}
K.~E.~Cahill, A.~Comtet and R.~J.~Glauber,
``Mass Formulas for Static Solitons,''
Phys. Lett. B \textbf{64} (1976), 283-285
doi:10.1016/0370-2693(76)90202-1

\bibitem{mekink}
J.~Evslin,
``Manifestly Finite Derivation of the Quantum Kink Mass,''
JHEP \textbf{11} (2019), 161
doi:10.1007/JHEP11(2019)161
[arXiv:1908.06710 [hep-th]].

\bibitem{mephi42}
J.~Evslin,
``$\phi^4$ kink mass at two loops,''
Phys. Rev. D \textbf{104} (2021) no.8, 085013
doi:10.1103/PhysRevD.104.085013
[arXiv:2104.07991 [hep-th]].

\bibitem{meunbind}
J.~Evslin,
``Evidence for the unbinding of the $\phi^{4}$ kink\textquoteright{}s shape mode,''
JHEP \textbf{09} (2021), 009
doi:10.1007/JHEP09(2021)009
[arXiv:2104.14387 [hep-th]].

% BPS-imp model

\bibitem{impurita}
C.~Adam, T.~Romanczukiewicz and A.~Wereszczynski,
``The $\phi^4$ model with the BPS preserving impurity,''
JHEP \textbf{03} (2019), 131
doi:10.1007/JHEP03(2019)131
[arXiv:1812.04007].



% BPS-imp other 

\bibitem{susy} C. Adam, J. Queiruga, A. Wereszczynski, ``BPS soliton-impurity models and supersymmetry", JHEP 1907 (2019) 164 [arXiv:1901.04501]. 

\bibitem{solvable} C. Adam, K. Oles, J. Queiruga, T. Romanczukiewicz, A. Wereszczynski, ``Solvable self-dual impurity models",  JHEP 1907 (2019) 150 [arXiv:1905.06080]. 

\bibitem{azadeh} J. Campos, A. Mohammadi, ``Fermion transfer in the $\phi^4$ model with a half-BPS preserving impurity", Phys. Rev. D 102 (2020) 045003. 

\bibitem{susy-2} C. Kim, Y.  Kim, O-Kab Kwon, ``Supersymmetric Inhomogeneous Field Theories in 1+1 Dimensions", JHEP 01 (2022) 140.  

\bibitem{wpol}
H.~Weigel,
``Vacuum Polarization Energy for General Backgrounds in One Space Dimension,''
Phys. Lett. B \textbf{766} (2017), 65-70
doi:10.1016/j.physletb.2016.12.055
[arXiv:1612.08641 [hep-th]].

\bibitem{tstabile}
T.~Roma\'nczukiewicz,
``Could the primordial radiation be responsible for vanishing of topological impuritys?,''
Phys. Lett. B \textbf{773} (2017), 295-299
doi:10.1016/j.physletb.2017.08.045
[arXiv:1706.05192 [hep-th]].

\bibitem{wstabile}
H.~Weigel,
``Quantum Instabilities of Solitons,''
AIP Conf. Proc. \textbf{2116} (2019) no.1, 170002
doi:10.1063/1.5114153
[arXiv:1907.10942 [hep-th]].

\bibitem{sw-2fields}  C. Adam, K. Oles, T. Romanczukiewicz, A. Wereszczynski,
and W. Zakrzewski, "Spectral walls in multifield kink dynamics", \textit{JHEP} \textbf{08} (2021) 147 [arXiv:2105.14771].

\bibitem{eg-sw-1} D. Bazeia, J.R.S. Nascimento, R.F. Ribeiro and D. Toledo, ``Soliton stability in systems of two real scalar fields", J. Phys. A 30 (1997) 8157 [hep-th/9705224].

\bibitem{eg-sw-2} A.A. Izquierdo, M.A.G. Leon and J.M. Guilarte, ``The kink variety in systems of two coupled scalar fields in two space-time dimensions", Phys. Rev. D 65 (2002) 085012 [hep-th/0201200].

\bibitem{eg-sw-3} D. Bazeia, L. Losano and C. Wotzasek, ``Domain walls in three field models", Phys. Rev. D 66 (2002) 105025 [hep-th/0206031].

\bibitem{eg-sw-4} A.A. Izquierdo, M.A. Gonzalez Leon, J.M. Guilarte and M. de la Torre Mayado, ``Adiabatic motion of two component BPS kinks", Phys. Rev. D 66 (2002) 105022 [hep-th/0207064].

\bibitem{eg-sw-5} L.A. Ferreira, P. Klimas and W.J. Zakrzewski, ``Self-dual sectors for scalar field theories in (1 + 1) dimensions", JHEP 01 (2019) 020 [arXiv:1808.10052].

\bibitem{eg-sw-6} L.A. Ferreira, P. Klimas, A. Wereszczynski and W.J. Zakrzewski, ``Some comments on BPS systems", J. Phys. A 52 (2019) 315201 [arXiv:1803.08985].


\bibitem{memassa}
J.~Evslin,
``Well-defined quantum soliton masses without supersymmetry,''
Phys. Rev. D \textbf{101} (2020) no.6, 065005
doi:10.1103/PhysRevD.101.065005
[arXiv:2002.12523 [hep-th]].

\bibitem{deglift}
A.~Alonso-Izquierdo and J.~M.~Guilarte,
``Quantum-induced interactions in the moduli space of degenerate BPS domain walls,''
JHEP \textbf{01} (2014), 125
doi:10.1007/JHEP01(2014)125
[arXiv:1307.0740 [hep-th]].

\bibitem{Thick}
	C. Adam, K. Oles, T. Romanczukiewicz and A. Wereszczynski,
	``Kink-antikink collisions in a weakly interacting $\phi^4$ model,"
	Phys. Rev. E \textbf{102} (2020), 062214
	[arXiv:hep-th/1912.09371].


\bibitem{horvath21}
D.~X.~Horv\'ath, M.~Kormos, S.~Sotiriadis and G.~Tak\'acs,
``Inhomogeneous quantum quenches in the sine-Gordon theory,''
[arXiv:2109.06869 [cond-mat.str-el]].

\bibitem{kovtun21}
A.~Kovtun,
``Analytical computation of quantum corrections to non-topological soliton (bright soliton) within the saddle-point approximation,''
[arXiv:2110.05222 [hep-th]].

\bibitem{kovtun18}
A.~Kovtun, E.~Nugaev and A.~Shkerin,
``Vibrational modes of Q-balls,''
Phys. Rev. D \textbf{98} (2018) no.9, 096016
doi:10.1103/PhysRevD.98.096016
[arXiv:1805.03518 [hep-th]].

\bibitem{bt}
C.~Barnes and N.~Turok,
``A Technique for Calculating Quantum Corrections to Solitons,''
[arXiv:hep-th/9711071 [hep-th]].

\bibitem{gh}
S.~B.~Gudnason and C.~Halcrow,
``Vibrational modes of Skyrmions,''
Phys. Rev. D \textbf{98} (2018), 125010
doi:10.1103/PhysRevD.98.125010
[arXiv:hep-th/1811.00562 [hep-th]].



\end{thebibliography}
\end{document}